\RequirePackage{xcolor}
\documentclass[journal,twoside,web]{ieeecolor}
\usepackage{generic}
\usepackage{booktabs}
\usepackage{tikz}
\usetikzlibrary{shapes,arrows, fit}
\usepackage{comment}
\usepackage{cite}
\usepackage{amssymb}
\usepackage{graphicx} % for pdf, bitmapped graphics files
\usepackage{epstopdf}
\usepackage{caption}
\usepackage{algpseudocode}
\usepackage{algorithm}
\usepackage{balance}
\usepackage[fancythm,fancybb]{jphmacros2e} 
\usepackage{amsfonts}
\usepackage{mathrsfs}
\usepackage{amsmath}
\usepackage{mathtools}
 \usepackage[norefs,nocites]{refcheck}
\let\labelindent\relax
\usepackage{enumitem}

\usepackage{subcaption}

\newcommand{\norm}[1]{\left\lVert#1\right\rVert}

\usepackage{titlesec}
\titlespacing{\section}{0pt}{0.5\baselineskip}{0.5\baselineskip}
\titlespacing{\subsection}{0pt}{0.4\baselineskip}{0.4\baselineskip}
\makeatletter
\g@addto@macro\normalsize{%
\setlength\abovedisplayskip{5pt}            %Equation space
\setlength\belowdisplayskip{5pt}
\setlength\abovedisplayshortskip{5pt}
\setlength\belowdisplayshortskip{5pt}
}
\makeatother

\tikzstyle{block} = [draw, fill=blue!20, rectangle, 
	minimum height=3em, minimum width=3em]
	\tikzstyle{joint} = [draw, fill=black, circle, 
	inner sep=0pt, minimum size=2pt]
	\tikzstyle{sum} = [draw, fill=blue!20, circle, node distance=1cm]
	\tikzstyle{input} = [coordinate]
	\tikzstyle{output} = [coordinate]
	\tikzstyle{pinstyle} = [pin edge={to-,thin,black}]

\begin{document}
%\IEEEoverridecommandlockouts
	\allowdisplaybreaks

\title{Deception Against Data-Driven Linear-Quadratic Control}
\author{Filippos Fotiadis$^1$,~\IEEEmembership{Member,~IEEE}, Aris Kanellopoulos$^2$,~\IEEEmembership{Member,~IEEE}, \\Kyriakos G. Vamvoudakis$^3$,~\IEEEmembership{Senior~Member,~IEEE}, Ufuk Topcu$^1$,~\IEEEmembership{Fellow,~IEEE}   
\thanks{$^1$ F. Fotiadis and U. Topcu are with the Oden Institute for Computational Engineering \& Sciences, University of Texas at Austin, Austin, TX, USA. Email:
\{ffotiadis, utopcu\}@utexas.edu.}
\thanks{$^2$ A. Kanellopoulos is with the  Division of Decision and Control Systems, School of Electrical Engineering and Computer Science, KTH Royal Institute of Technology, Stockholm,
Sweden, e-mail: arisk@kth.se.}
\thanks{$^3$K. G. Vamvoudakis is with the School of Aerospace Engineering, 
Georgia Institute of Technology, Atlanta, GA, USA. Email:
kyriakos@gatech.edu.}
\thanks{
This work was supported in part by ARO under grant Nos. W911NF-24-1-0174 and W911NF-23-1-0317, by ONR under grant No. N00014-22-1-2703, by NASA under grant No. 80NSSC24M0070, and by NSF under grant Nos. CAREER CPS-1851588,  SLES-2415479,  SATC-2231651, CPS-2227185, CPS-2227153.}
}
\maketitle

\begin{abstract}
    Deception is a common defense mechanism against adversaries with an information disadvantage. It can force such adversaries to select suboptimal policies for a defender's benefit.  We consider a setting where an adversary tries to learn the optimal linear-quadratic attack against a system, the dynamics of which it does not know. On the other end, a defender who knows its dynamics exploits its information advantage and injects a deceptive input into the system to mislead the adversary.  The defender's aim is to then strategically design this deceptive input: it should force the adversary to learn, as closely as possible, a pre-selected attack that is different from the optimal one.  We show that this deception design problem boils down to a solution of
    a coupled algebraic Riccati and a Lyapunov equation which, however, are challenging to tackle analytically. Nevertheless, we use a block successive over-relaxation algorithm to extract their solution numerically and prove the algorithm's convergence under certain conditions. 
    We perform simulations on a benchmark aircraft, where we showcase how the proposed algorithm can mislead adversaries into learning attacks that are less performance-degrading.
\end{abstract}


\section{Introduction}

Deception is a mechanism defenders often use
in adversarial environments, particularly as a tool to mislead their opponents into adopting oblivious policies. For example, adversarial entities can employ deception in cyber-physical settings as a means to render their attacks unnoticed \cite{hespanha2019output, bai2017kalman},
but also, in other contexts such as robotics \cite{shim2014robot}, autonomous vehicles \cite{mceneaney2005deception}, and warfare \cite{whaley2007stratagem}. Despite the heterogeneity of these applications, a common theme among them is that of information asymmetry; that is, to be able to deceive the opponent, one must have an information advantage against them, and the opponent should usually be oblivious to this advantage.

From a control-theoretic perspective, deception in dynamical systems can be formulated within a broader optimal control and game-theoretic framework. In such settings, one agent strategically perturbs information or system inputs in order to influence the policy computed or learned by one or more other agents \cite{vamvoudakis2025deception}. This leads naturally to bilevel optimal control problems in which one decision maker anticipates and shapes the optimization or learning process of another.

In this paper, we consider deception in the context of data-driven linear-quadratic control. This is a recently emerged field in control theory that aims to design optimal controllers for systems 
whose dynamics are either uncertain or completely unknown. Its underlying idea is to use state and/or input histories of the system at hand to synthesize the desired control policy, instead of solving model-based equations that require strong prior empirical knowledge \cite{de2019formulas, lewis2009reinforcement}.
Yet, a major backdoor in data-driven control is its sensitivity to perturbations often present in the measured data. Such perturbations can be benign when they conform to specific random patterns or are sufficiently small \cite{berberich2020data, zhongpingrobust}. However, this is not the case when they are specifically crafted by an opponent with an information advantage as a means for deception. Crafting such deception is the focus of this paper.

More specifically, we consider a setting where an adversary wants to compute the optimal linear quadratic attack against a dynamical system. However, being an outsider, the adversary has no prior knowledge regarding the system's dynamics. It thus needs to either perform a direct data-driven method to learn the optimal attack or employ system identification and then learn the optimal attack indirectly. On the other end, the defender of the system is aware of the adversary's malicious objective and is motivated to perform deception to mislead them towards learning a more benign attack. To that end, the defender exploits its information advantage with respect to the system's dynamics and injects deceptive feedback into the system to distort the data the adversary is using. A key characteristic of such feedback is that it has the same effect on the attack that the adversary will ultimately learn, irrespective of the adversary's actual learning algorithm.

To obtain the defender's deceptive input, we cast its objective as a constrained optimization problem balancing two objectives: i) the desire to steer the adversary's learned policy towards an a priori chosen benign gain; and ii) the desire to use as small of deceptive feedback as possible to maintain the nominal closed-loop characteristics and stability. We mathematically specify the conditions under which the defender's  optimization is well-defined and characterize its solution as the root of a coupled algebraic Riccati equation and a Lyapunov equation. Subsequently, since these coupled equations are difficult to analytically solve, we 
  tackle them numerically with a successive over-relaxation iterative algorithm. In addition, we prove this algorithm's convergence under certain conditions. We finally note that the proposed deception design also applies to the easier problem of deceiving minimizing data-driven linear-quadratic regulators.

\subsection*{Related work}

Despite data-driven control being a relatively young field, relevant studies regarding its deception have recently appeared in the literature.
For example, related to the philosophy of this paper is the concept of deception in data-driven Nash equilibrium seeking \cite{tang2024deception}. It considers a setting where an agent engaged in a multi-player game has an information advantage and uses this advantage to design a deceptive input that steers the other agents' policies toward a deceptive equilibrium. Nevertheless, the setting considered in \cite{tang2024deception} is tailored to repeated static games. Related Nash equilibrium seeking problems under uncertainty and adversarial disturbances have also been studied in \cite{Feng2026TAC,Feng2026JAS}, though these works focus on convergence to the true equilibrium rather than intentionally steering the learning process toward a prescribed policy. 

The concept of deceiving data-driven methods is also often related to poisoning such methods, owing to the data corruption that implicitly or explicitly takes place as the means for deception. This idea of poisoning has attracted much interest in the machine learning community \cite{munoz2017towards, jagielski2018manipulating, biggio2012poisoning, chen2017targeted, shafahi2018poison} and a few studies have also investigated it from a systems and control theoretic perspective. For instance, relevant examples include poisoning virtual reference feedback controllers \cite{russo2021data, russo2021poisoning} and data predictive control \cite{yu2023poisoning}. Nevertheless, most of these techniques revolve around distantly related setups with discrete-time systems, different objectives, and sometimes without theoretical convergence guarantees. Relevant research initiatives have also focused on increasing resilience against poisoning attacks \cite{mao2022decentralized, khazraei2022resiliency, zhang2017game, chekan2023regret}. However, the problem of explicitly deceiving a learning-based optimal control algorithm, so that it converges to a prespecified policy, has not been studied.

Moving target defense is another mechanism used to deal with adversaries \cite{kanellopoulos2019moving, sandbergmoving, sinopolimoving}. Its main idea is that to prevent an adversary from stealthily attacking a dynamical system, it is in the interest of the defender to constantly alter the system's dynamics to reveal or hinder the attack. However, moving target defense does not exploit the defender's information advantage to explicitly deceive the adversary's policy. This is the case in the deception method we develop, where one can prescribe how the defender's deceptive input will affect the learned adversarial policy and hence steer it towards an a priori chosen benign target. While worst-case min-max approaches as in \cite{bacsar1998dynamic, attacks1, attacks2} aim to protect against fully-informed adversaries, we exploit the defender's advantage to strategically control the adversary's policy. 

A preliminary version of this paper appeared in \cite{fotiadis2024poisoning}, which studied the problem of deception against minimizing data-driven linear quadratic regulators. However, \cite{fotiadis2024poisoning} only provided simulation results to support its algorithm's ability in extracting the optimal deception policy. On the other hand, hereon, we adopt a stricter approach and theoretically prove the algorithm's convergence properties. In addition, we shed light on the problem of deceiving  maximizing  regulators. This is a much more challenging problem owing to the associated value function possibly becoming unbounded and thus undefined \cite{bacsar1998dynamic}.

The contributions of the paper are summarized as follows.
\begin{enumerate}
\item We consider deception in a setting where a defender, who has knowledge of the system's matrices, injects an additional state-feedback input to strategically influence the learning dynamics of a data-driven attacker, with the goal of steering it toward a suboptimal attack policy. Unlike classical LQR game formulations \cite{bacsar1998dynamic, attacks1, attacks2}, where all players have model access and the equilibrium is uniquely determined by the problem parameters, our setting exploits the fact that the adversary must learn its policy from data, opening the door to shaping what it converges to. This also distinguishes our approach from moving target defense \cite{kanellopoulos2019moving, sandbergmoving, sinopolimoving}, which does not explicitly target the adversary’s learned policy, and from deception in Nash equilibrium seeking \cite{tang2024deception}, which considers static game settings.
\item We characterize the optimal deception gain as a solution of a coupled algebraic Riccati and Lyapunov equation, and develop a block successive over-relaxation algorithm with provable convergence guarantees. This provides theoretical foundations for this specific problem that are absent from related data poisoning approaches, such as those on virtual reference feedback controllers \cite{russo2021data, russo2021poisoning} and data predictive control \cite{yu2023poisoning}, as well as from the preliminary version of this paper \cite{fotiadis2024poisoning}.
\item We analyze robustness to uncertainty in the adversary's cost matrices and validate the framework on an aircraft model, demonstrating substantial attack suppression.
\end{enumerate}

\textit{Structure:} The remainder of the paper is organized as follows. 
Section~\ref{sec:prel} presents preliminaries, Section~\ref{sec:pr} formulates the deception problem, and Section~\ref{sec:design} outlines the proposed solution within an algorithmic framework. 
Section~\ref{sec:conv} establishes convergence guarantees for the algorithm, while Section~\ref{sec:discuss} discusses robustness to parametric uncertainty and relaxed assumptions. 
Finally, Section~\ref{sec:sim} presents simulation results, and Section~\ref{sec:conc} concludes the paper.

\section{Preliminaries}\label{sec:prel}

\textit{Notation:} We denote as $\mathbb{R}$ the set of real numbers, and as $\mathbb{N}$ the set of natural numbers including zero. 
The vector $e_i\in\mathbb{R}^n$ denotes a unit vector with a unity located in the $i$-th row.
For any matrix $X$, $\norm{X}_{\mathrm{F}}$ denotes its Frobenius norm,  whereas $\mathrm{Ran}(X)$ denotes its range (i.e., its column space). In addition, $[X]_{ij}$ denotes the entry in its $i$-th row and $j$-th column. Moreover, we follow the matrix differentiation convention $\left[\frac{\mathrm{d}}{\mathrm{d}X}\right]_{ij}=\frac{\mathrm{d}}{\mathrm{d}[X]_{ij}}$. For a square matrix $X$, $\alpha(X)$ denotes its spectral abscissa, and $\textrm{tr}(X)$ its trace. For a symmetric matrix $X$, $\lambda_{\mathrm{min}}(X)$ and $\lambda_{\mathrm{max}}(X)$ denote its minimum and maximum eigenvalue, respectively. We denote $X\succ0$ (respectively, $X\succeq0)$ if $X$ is positive definite (respectively, positive semidefinite). We denote as $\textrm{vec}(X)$ the vectorization of $X$. We use $I_n$ to denote the identity matrix of order $n$  (we omit the subscript when the order is obvious). 
We denote the Kronecker delta function as $\delta_{i,j}$, so that $\delta_{i,j}=1$ if $i=j$, and $\delta_{i,j}=0$ else. 
\begin{definition}

Consider the algebraic Riccati equation:
\begin{equation}\label{eq:ARE_notation}
A^\mathrm{T}P+PA+Q-P(B_1R_1^{-1}B_1^\mathrm{T}-B_2R_2^{-1}B_2^\mathrm{T})P=0,
\end{equation}
where $A,P,Q\in\mathbb{R}^{n\times n}$, $R_1\in\mathbb{R}^{m_u\times m_u}$, $R_2\in\mathbb{R}^{m_a\times m_a}$, $B_1\in\mathbb{R}^{n\times m_u}$, $B_2\in\mathbb{R}^{n\times m_a}$,  $R_1,R_2,Q\succ0$, and $(A,B_1)$ is stabilizable.
When it exists, we denote the unique stabilizing solution of \eqref{eq:ARE_notation} as $P\in\mathbb{S}_+$,  with  $\mathbb{S}_+$ being the set of positive-definite stabilizing solutions. Such a solution is positive definite, though not necessarily uniquely, but is the minimal such solution of \eqref{eq:ARE_notation} \cite{bacsar1998dynamic}. 
\end{definition}

\section{Problem Formulation}\label{sec:pr}

\subsection{Closed-Loop System}

Consider the continuous-time, closed-loop system for all $t\geq0$:
\begin{equation}\label{eq:sys}
\dot{x}(t)=Ax(t),
\end{equation}
where $x(t)\in\mathbb{R}^n$ is the state of the system, and $A\in\mathbb{R}^{n\times n}$ is the closed-loop state matrix.  By closed loop, we mean that
$A=A_o+B_uF$,
where $A_o\in\mathbb{R}^{n\times n}$ is the open-loop state matrix, $B_u\in\mathbb{R}^{n\times m_u}$ is the input matrix, and $F\in\mathbb{R}^{m_u\times n}$ is a stabilizing gain such that $A$ is Hurwitz. For example, $F$ could be a gain obtained from a pole-placement procedure that moves the eigenvalues of $A$ to the open left-half plane. For such a gain to be feasible, $(A, B_u)$ must also be controllable. 

\subsection{The Adversary's Model and Objectives}

We assume that an adversary can launch an adversarial attack on system \eqref{eq:sys}. In the presence of such an attack, system \eqref{eq:sys} would take the form
\begin{equation}\label{eq:sysa}
\dot{x}(t)=Ax(t)+B_aa(t),
\end{equation}
with $B_a\in\mathbb{R}^{n\times m_a}$ being the adversary's input matrix, and $a(t)\in\mathbb{R}^{m_a}$ being the attack. The adversary's purpose is to then design the attack to balance two objectives: i) disturbing the stabilization of \eqref{eq:sysa}; and ii) moderating the attack's magnitude. We capture this dual objective with the maximizing linear-quadratic control problem:
\begin{align}\nonumber
     \max_{K\in\mathbb{R}^{ m_a\times n}} ~&\int_0^\infty(x^\mathrm{T}(\tau)Qx(\tau)-a^\mathrm{T}(\tau)Ra(\tau))\mathrm{d}\tau,\\ 
     \textrm{s.t.} \qquad &\dot{x}(t)=Ax(t)+B_aa(t),\label{eq:Jnom}\\
    & a(t)=Kx(t),\nonumber
\end{align}
where $Q,~R\succ0$ are weighting matrices and $K$ is the adversary's attack gain that is to be optimized. Provided that $\lambda_{\textrm{min}}(R)$ is above a threshold, it then follows from standard linear systems theory \cite{bacsar1998dynamic} that the solution to this problem is given by
\begin{equation}\label{eq:K}
K^\star=R^{-1}B_a^\mathrm{T}P,
\end{equation}
where $P\in\mathbb{S}_+$ is the minimal positive definite solution of the algebraic Riccati equation (ARE)
\begin{equation}\label{eq:are}
A^\mathrm{T}P+PA+Q+PB_aR^{-1}B_a^\mathrm{T}P=0.
\end{equation}

We now present the information structure of the adversary.
\begin{assumption}
The following hold true.
\begin{enumerate}
\item The adversary does not know the values of the entries of the system matrix $A$.   They may or may not know the values of the entries of the system matrix $B_a$. 
\item The adversary can measure the state of the system. 
\end{enumerate}
\end{assumption}
The main implication of the adversary's information structure is that it impedes them from directly solving the ARE \eqref{eq:are}, and hence from obtaining the optimal attack gain $K^\star$ in \eqref{eq:K}. In that regard, the adversary is forced to gather measurements from the system over a given window and use them to either solve \eqref{eq:Jnom} in a direct data-driven manner, or to perform system identification and then solve \eqref{eq:Jnom}. 

\subsection{The Defender's Model and Objectives}

We assume that the defender of the system is aware that a potential adversary might be gathering information from the system with the aim of synthesizing the optimal attack gain $K^\star$ in \eqref{eq:K}.   In response, the defender may temporarily modify the system’s dynamics to induce the adversary to learn a prespecified gain different from $K^\star$.   Specifically, during this window, the system's dynamics take the form
\begin{equation}\label{eq:sysua}
\dot{x}(t)=Ax(t)+B_uu(t)+B_aa(t),
\end{equation}
where $u(t)\in\mathbb{R}^{ m_u}$ is the defender's deceiving input added to spoof the system dynamics and deceive the adversary, forcing them to learn a gain $\bar{K}\in\mathbb{R}^{m_a\times n}$ that is different from $K^\star$. 

We now present the information structure of the defender.
\begin{assumption}\label{ass:2}
The following hold true.
\begin{enumerate}
\item The defender knows the system matrices $A, B_u, B_a$.
\item The defender can measure the state of the system.
\item The defender does not know the learning algorithm the adversary is using, but knows its objective. 
\end{enumerate}
\end{assumption}

\begin{figure}[!t]
		\begin{center}

			\begin{tikzpicture}[auto, node distance=2cm,>=latex']
			
			\node [input, name=a] {};
			\node [block, right of=a,
			node distance=3.5cm,    align=center] (sys) {Closed-Loop System};
            \node [output, name=x, right of=a,
			node distance=6.9cm] {};
            \node [block, name=L, below of=sys,
			node distance=1.5cm] {$\Lambda$};
			
        \node at (1.1,1.0) (tmp1) {};
			\node at (5.9,-2.2) (tmp2) {};
			
             \fill[red, fill opacity=0.1, draw=red] (tmp1) rectangle (tmp2);
			\node at (3.5,1.2) (tmp3) {Deceptive System};

            \node [joint, right of=sys, node distance=2cm] (jointx) {};
            \draw (jointx) -- ++(0,-1.5) [->] ->  (L);
            
			\draw [->] (a) -- node {\hspace{-8mm}$a(t)~$} (sys);
			\draw [->] (sys) -- node {\hspace{8mm}$x(t)~$} (x);

        \draw (L)  -- node {$u(t)~$} ++(-2.0,0) [->] -| ++(0,1) [->] |- node {} ([yshift=-4mm]sys);
			\end{tikzpicture}
			\caption{The deception scheme. The adversary uses state data $x(t)$ and input data $a(t)$ to learn the optimal attack \eqref{eq:K} against the closed-loop system \eqref{eq:sysa}. However, during the adversary's data-gathering process, the defender feeds into the closed loop a deceptive gain $\Lambda$, forcing the adversary to learn the incorrect attack \eqref{eq:KL} instead. }  \label{fig:structure}
		\end{center} 
	\end{figure}
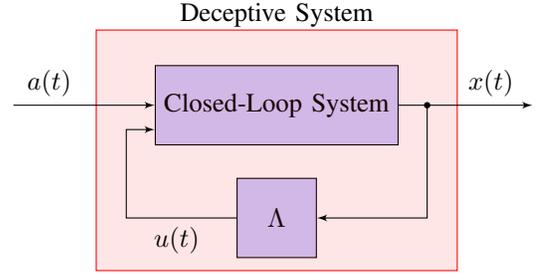

The fact that the defender does not know the learning algorithm of the adversary, though realistic, makes it difficult to gauge how the deceiving input $u(t)$ will affect where that algorithm will converge. An exception is if the defender chooses $u$ as
\begin{equation*}
u(t)=\Lambda x(t),
\end{equation*}
 where $\Lambda\in \mathbb{R}^{ m_u\times n}$ is an appropriately chosen deception gain. Such a strategy simply redefines the system's internal dynamics, and hence will generate the same effect to the learned adversarial gain irrespective of the underlying learning dynamics. 
Hence, if the adversary employs a learning-based scheme to estimate the optimal gain $K^\star$ that solves \eqref{eq:Jnom} while under the proposed deceptive measure, the adversary will instead learn the gain that solves
\begin{align}\nonumber
     \max_{K\in\mathbb{R}^{m_a\times n}} ~&\int_0^\infty(x^\mathrm{T}(\tau)Qx(\tau)-a^\mathrm{T}(\tau)Ra(\tau))\mathrm{d}\tau,\\ 
     \textrm{s.t.} \qquad &\dot{x}(t)=(A+B_u\Lambda)x(t)+B_aa(t),\label{eq:Jper}\\
     &a(t)=Kx(t),\nonumber
\end{align}
where we note that the trajectories in this maximization problem are over the spoofed dynamics, whereas the trajectories in \eqref{eq:Jnom} were over the dynamics of the original system.  Following again standard linear systems theory, one can show that the optimal solution to \eqref{eq:Jper} is
\begin{equation}\label{eq:KL}
    K_u(\Lambda)=R^{-1}B_a^\mathrm{T}P_u(\Lambda),
\end{equation}
where $P_u(\Lambda)\in\mathbb{R}^{n\times n}\in\mathbb{S}_+$ is now the minimal positive definite solution of the ``spoofed" ARE:
\begin{multline}\label{eq:RicL}
    (A+B_u\Lambda)^\mathrm{T}P_u+P_u(A+B_u\Lambda)+Q\\+P_uB_aR^{-1}B_a^\mathrm{T}P_u=0.
\end{multline} 
 Figure \ref{fig:structure} depicts this deceptive process. 

The above equations imply that the defender can predict the adversarial gain $K_u(\Lambda)$ that the enemy will ultimately learn, particularly by exploiting its information advantage with respect to the system's dynamics.  In addition, the defender can perform this prediction without using any data. Nevertheless, choosing the deception gain $\Lambda$ to control $K_u(\Lambda)$ is far from being a trivial task, because the dependence of $K_u(\Lambda)$ on $\Lambda$ goes through the minimal solution of the ARE \eqref{eq:RicL}. In that regard, if the defender wants, for example, to suppress the adversarial attack by forcing $K_u(\Lambda)$ to be very close to zero, then it is not clear how it should select $\Lambda$ to achieve this suppression optimally.
Motivated by this issue, we formalize the problem we investigate in the remainder of the paper as follows. 

\begin{problem}\label{pr:1}
    Design $\Lambda$, so that the gain $K_u(\Lambda)$ that the adversary will learn closely resembles a different gain $\bar{K}\in\mathbb{R}^{m_a\times n}$ that is prechosen by the defender.
\end{problem}

\begin{remark}
Another variation of the deception problem considered in this paper is one where a defender tries to learn the  minimizing linear-quadratic regulator for the system, and an adversary tries to deceive them towards converging to an unrelated, suboptimal gain. We note that this problem is easier owing to its inherently convex nature, and one can solve it using tools identical to those presented in the following sections; see Section \ref{sec:minimizing} for more details. 
\end{remark}

\begin{remark}
While we assume the defender has full state feedback, the approach can also be extended to the output feedback case through the use of an observer or estimator.  However, estimation errors might affect the impact of deception.  
\end{remark}

\section{Deception Design}\label{sec:design}

\subsection{Deception Design using a Distance-Based Metric}

One method the defender could use to design its deception gain $\Lambda$, is so that $K_u(\Lambda)$ is as close as possible---in the matrix norm sense---to $\bar{K}$.  In  other words, the defender could select its gain $\Lambda$ so that it minimizes  $\norm{K_u(\Lambda)-\bar{K}}_\mathrm{F}^2$. Under this objective, 
the optimal deception gain would be derived as the solution of the constrained optimization
\begin{align}\nonumber
&\inf_{\Lambda\in\mathbb{R}^{m_u\times n},P_u\in\mathbb{R}^{n\times n}}J(\Lambda,P_u)=\norm{K_u(\Lambda)-\bar{K}}_{\mathrm{F}}^2+\mathrm{tr}(\Lambda^\mathrm{T}\Gamma\Lambda),\\
&~\quad\textrm{s.t.} \qquad\quad \eqref{eq:KL}, \eqref{eq:RicL}, \label{eq:opt1}\\
&\qquad\qquad\quad~ P_u\in\mathbb{S}_+, \nonumber
\end{align}
where $\bar{K}$ is as in Problem \ref{pr:1}, $ \Gamma\succ0$, and $\mathrm{tr}(\Lambda^\mathrm{T}\Gamma\Lambda)$ is a regulation term. This regulation term is necessary to ensure that the minimum exists, but also to penalize how large the deception gain $\Lambda$ can be to 
minimize  deviation from nominal closed-loop performance  and maintain stability\footnote{This is for the case where $\bar{K}$ is not stabilizing. For the case where $\bar{K}=0$, regularization is not needed; see Section \ref{sec:discuss1}. }. As an alternative, the distance of the target gain $\bar{K}$ from the original gain $K^\star$ should be sufficiently small,  assuming $K^\star$ exists. We provide an explicit -- though conservative -- condition on the distance $\norm{K^\star-\bar{K}}_{\mathrm{F}}$ as well as the regulation weight $\Gamma$, for ensuring closed-loop stability and the feasibility of \eqref{eq:opt1}, in the following Lemma. The Lemma's proof, as well as the proofs of all technical results, are postponed to the Appendix.

\begin{lemma}\label{le:stab}
Let $S\in\mathbb{R}^{n\times n}$ be the unique positive-definite solution of
\begin{equation}\label{eq:LE_S}
(A+B_aK^\star)^\mathrm{T}S+S(A+B_aK^\star)+2I=0.
\end{equation}
Choose $\Gamma$ and the target gain $\bar{K}$ so that
\begin{equation}\label{eq:stab_condition}
\norm{K^\star-\bar{K}}_{\mathrm{F}}< \frac{\sqrt{\lambda_{\textrm{min}}(\Gamma)}}{\sqrt{\lambda_{\textrm{min}}(\Gamma)}\norm{SB_a}_{\mathrm{F}}+\norm{SB_u}_{\mathrm{F}}}.
\end{equation}
Then, \eqref{eq:opt1} admits a global minimizer $(\Lambda^\star,P_u^\star)$, i.e., its infimum is a minimum. In addition, 
$A+B_u\Lambda^\star+B_aK_u(\Lambda^\star)$ is strictly stable. 
\end{lemma}

We note that the condition of Lemma \ref{le:stab}, though sufficient, is not necessary for maintaining stability and feasibility. It 
can often prove to be quite conservative, and merely having a sufficiently large regulation matrix $\Gamma$ in terms of eigenvalues should suffice both in practice and theory. Moreover, 
simulation results and an additional analysis in Section \ref{sec:discuss2} show that even this condition is often redundant, as for certain reasonable choices of $\bar{K}$ the regulation matrix $\Gamma$ can be arbitrarily small; see Section \ref{sec:discuss1} for further details.

To proceed, we use the Karush–Kuhn–Tucker conditions to characterize the solution to the optimization problem \eqref{eq:opt1}. 
\begin{theorem}\label{th:1}
Let  $(\Lambda^\star, P_u^\star)\in\mathbb{R}^{m_u\times n}\times\mathbb{R}^{n\times n}$ be the optimal solution to \eqref{eq:opt1}. Then, $P_u^\star\in\mathbb{S}_+$, and there exists $\Pi^\star\in\mathbb{R}^{n\times n}$ symmetric such that $(\Lambda, P_u, \Pi)=(\Lambda^\star, P_u^\star, \Pi^\star)$  are a solution to the system of equations
 \begin{subequations}
\begin{align}\nonumber
&(A+B_u\Lambda)^\mathrm{T}P_u+P_u(A+B_u\Lambda)\\&\qquad\qquad\qquad\qquad+Q+P_uB_aR^{-1}B_a^\mathrm{T}P_u=0,\label{eq:eq1}\\\nonumber
&\nonumber(A+B_u\Lambda+B_aR^{-1}B_a^\mathrm{T}P_u)\Pi\\&\nonumber+\Pi(A+B_u\Lambda{+}B_aR^{-1}B_a^\mathrm{T}P_u)^\mathrm{T}-\bar{K}^\mathrm{T}R^{-1}B_a^\mathrm{T}\\&-B_aR^{-1}\bar{K}+P_uB_aR^{-2}B_a^\mathrm{T}+B_aR^{-2}B_a^\mathrm{T}P_u=0,\label{eq:eq2}\\
&\Lambda=-\Gamma^{-1}B_u^\mathrm{T}P_u\Pi.\label{eq:eq3}
\end{align}
\end{subequations}
\end{theorem}

According to Theorem \ref{th:1}, we can extract
the optimal solution to the deception design problem \eqref{eq:opt1}  as a solution of the matrix equations \eqref{eq:eq1}-\eqref{eq:eq3}. 
However, a close inspection of these equations reveals that they are not easy to solve analytically, even in the scalar case. This is mainly due to their coupled and nonlinear nature, arising from the cross-term $P_uB_u\Lambda$ in \eqref{eq:eq1}, and the cross-terms  $P_u\Pi$ and $\Lambda\Pi$ in \eqref{eq:eq2}. 

A potential solution to the aforementioned bottleneck
could be to employ off-the-shelf numerical solvers for nonlinear systems of equations on \eqref{eq:eq1}-\eqref{eq:eq3}, combined with standard heuristics that may aid their chances of success. However, not only would such a choice be computationally expensive since the number of unknowns scales quadratically with $n$ in \eqref{eq:eq1}-\eqref{eq:eq3}, but it would also have no theoretical guarantees of success. A more thoughtful numerical approach is thus clearly required.

\subsection{Proposed Numerical Solution}

To be able to numerically attack equations \eqref{eq:eq1}-\eqref{eq:eq3} in an efficient manner, it is important to first understand their structure. Even though as a whole this system of equations does not seem to fall within any known class of systems, each of its equations -- when viewed independently -- has a relatively familiar structure. More specifically, equation \eqref{eq:eq1} is an ARE when viewed with $P_u$ being its sole unknown variable, and such AREs are straightforward to solve for their minimal positive definite solution. In addition, equation \eqref{eq:eq2} is a Lyapunov equation (LE) when viewed with $\Pi$ being its sole unknown variable, and hence has a unique solution  -- as long as $P_u$ is minimal -- that we can extract even analytically. Finally, provided that both $P_u$ and $\Pi$ are known a priori, \eqref{eq:eq3} is a very simple computation that yields the deception gain $\Lambda$.

\begin{algorithm}[!t]
\caption{ Block Successive Over-relaxation Algorithm to solve \eqref{eq:eq1}-\eqref{eq:eq3}}
\hspace*{0.7em}\textbf{Input}: \parbox[t]{\dimexpr\linewidth-4.5em}{Tolerance $\epsilon>0$, step size $\omega\in (0,2)$, regulation weight $\Gamma\succ0$, desired gain $\bar{K}$, parameters $A,B_u,B_a,R,Q$.\strut}\\
\hspace*{\algorithmicindent} \textbf{Output}: \parbox[t]{\dimexpr\linewidth-1.5em}{Estimated solution $(\hat{\Lambda}, \hat{P}_u, \hat{\Pi})$ to \eqref{eq:eq1}-\eqref{eq:eq3}.\strut}
\begin{algorithmic}[1]
\Procedure{}{}
\State  Initialize $\Lambda^0$ so that $J(\Lambda^0,P_u(\Lambda^0))\le J(0,P_u(0))$. 
\State $i \leftarrow 0$.
\While{$i=0$ or $\norm{\Lambda^i-\Lambda^{i-1}}_{\mathrm{F}}\ge\epsilon\cdot\omega$}
\State Solve for $P_u^{\textrm{GS}}\in\mathbb{S}_+$ from the ARE
\begin{multline}\label{eq:it1}
(A+B_u\Lambda^i)^\mathrm{T}P_u^{\textrm{GS}}+P_u^{\textrm{GS}}(A+B_u\Lambda^i)\\+Q+P_u^{\textrm{GS}}B_aR^{-1}B_a^\mathrm{T}P_u^{\textrm{GS}}=0.
\end{multline}
\State Set $P_u^{i+1}\leftarrow P_u^{\textrm{GS}}$.
\State Solve for $\Pi^{\textrm{GS}}$ from the LE
\begin{align}
&\nonumber(A+B_u\Lambda^{i}+B_aR^{-1}B_a^\mathrm{T}P_u^{i+1})\Pi^{\textrm{GS}}\\&\label{eq:it2}{+}\Pi^{\textrm{GS}}(A{+}B_u\Lambda^{i}{+}B_aR^{-1}B_a^\mathrm{T}P_u^{i+1})^\mathrm{T}{-}\bar{K}^\mathrm{T}R^{-1}B_a^\mathrm{T}\\&-B_aR^{-1}\bar{K}+P_u^{i+1}B_aR^{-2}B_a^\mathrm{T}+B_aR^{-2}B_a^\mathrm{T}P_u^{i+1}=0.\nonumber
\end{align}
\State Set $\Pi^{i+1}\leftarrow \Pi^{\textrm{GS}}$.
\State Compute $\Lambda^{\textrm{GS}}$ as
\begin{equation}\label{eq:it3}
\Lambda^{\textrm{GS}}=-\Gamma^{-1}B_u^\mathrm{T}P_u^{i+1}\Pi^{i+1}.
\end{equation}
\State Set $\Lambda^{i+1}\leftarrow \Lambda^{i}+\omega(\Lambda^{\textrm{GS}}-\Lambda^{i})$.
\State $i\leftarrow i+1$.
\EndWhile
\State $\hat{\Lambda}\leftarrow \Lambda^i$,  $\hat{P}_u\leftarrow P_u^i$, $\hat{\Pi}\leftarrow \Pi^i$.
\EndProcedure
\end{algorithmic}\label{al:BSOR}
\end{algorithm}

In view of the above, a potential structured numerical method to solve \eqref{eq:eq1}-\eqref{eq:eq3} is the following: one can proceed to begin with an arbitrary gain $\Lambda$, then solve \eqref{eq:eq1} for its minimal positive definite solution $P_u$ (provided it exists), then solve \eqref{eq:eq2} for its symmetric solution (provided it exists), then update $\Lambda$ according to \eqref{eq:eq3}, and then repeat this process iteratively until a fixed point is found. Such a fixed point will necessarily be a solution to \eqref{eq:eq1}-\eqref{eq:eq3}, and hence qualify for being a stationary point of \eqref{eq:opt1}. 

The aforementioned method is known as the Gauss-Seidel method in numerical linear algebra \cite{ortega2000iterative}. We present a variation of it in Algorithm \ref{al:BSOR}, known as block successive over-relaxation, wherein the update of the gain $\Lambda$ at each iteration goes through a low-pass filter,  and where the iterates \eqref{eq:it1}-\eqref{eq:it3} imitate the structure of equations \eqref{eq:eq1}-\eqref{eq:eq3}. Unlike the Gauss-Seidel and the Jacobi method (Chapter 7 in \cite{ortega2000iterative}), which may diverge depending on the problem's parameters, successive over-relaxation has provable convergence guarantees as we show in Section \ref{sec:conv}. 

 \begin{remark}As further detailed in Section \ref{sec:pr},
the execution of Algorithm \ref{al:BSOR} does not require any data. The reason is that the defender has knowledge of the system's dynamics, and can use these to directly predict how the deceptive gain will affect the policy the attacker will learn. \end{remark}

\section{Convergence Analysis}\label{sec:conv}

We first presented Algorithm \ref{al:BSOR} in the preliminary study \cite{fotiadis2024poisoning}, which designed poisoning actuation attacks against defenders learning the minimizing linear-quadratic control of an unknown system. While simulation results in \cite{fotiadis2024poisoning} showcased that   Algorithm \ref{al:BSOR} performs well in a variety of settings, it is difficult to use such empirical results to argue about convergence properties, mainly because fixed-point iteration schemes are notorious for often being divergent. For this reason, 
 here we will adopt a theoretically strict approach, and mathematically prove that 
 Algorithm \ref{al:BSOR} leads to a stationary point of the deception problem \eqref{eq:opt1}. In addition, we will show that this stationary point, under certain conditions, corresponds to a minimum.

Towards this end, notice that in the optimization problem \eqref{eq:opt1}, the
ARE constraint \eqref{eq:RicL} completely dictates the choice of the matrix $P_u$. That is, once we select a specific gain $\Lambda$, then we subsequently obtain the matrix $P_u$ by solving the equality constraint \eqref{eq:RicL} for its minimal positive definite solution. Taking this into consideration, it is possible to define the cost function of \eqref{eq:opt1} uniquely with respect to $\Lambda$ as
\begin{equation}\label{eq:JL}
\tilde{J}(\Lambda)=J(\Lambda,P_u(\Lambda)),
\end{equation}
where, with a slight abuse of notation, $P_u(\cdot)$ is now a function with unknown analytical form, which maps $\Lambda$ to the minimal positive definite solution of the ARE
\begin{multline}\label{eq:PL}
(A+B_u\Lambda)^\mathrm{T}P_u(\Lambda)+P_u(\Lambda)(A+B_u\Lambda)\\+Q+P_u(\Lambda)B_aR^{-1}B_a^\mathrm{T}P_u(\Lambda)=0.
\end{multline}
In light of this definition, in what follows we prove the following central result: one can interpret the fixed point iteration presented in Algorithm \ref{al:BSOR} as a gradient iteration on the cost function $\tilde{J}(\Lambda)$, scaled by $\Gamma^{-1}$.

\begin{lemma}\label{le:grad}
For any $\Lambda\in \mathbb{R}^{m_u\times n}$ for which the ARE \eqref{eq:PL} admits a solution $P_u(\Lambda)\in\mathbb{S}_+$, we have
\begin{equation}\label{eq:grad_formula}
\frac{\mathrm{d}\tilde{J}(\Lambda)}{\mathrm{d}\Lambda}=2\left(\Gamma\Lambda+B_u^\mathrm{T}P_u(\Lambda)\Pi(\Lambda)\right),
\end{equation}
where $\Pi(\Lambda)$ is a matrix function of $\Lambda$, and the unique symmetric solution of the LE
\begin{align}\nonumber
&(A+B_u\Lambda+B_aR^{-1}B_a^\mathrm{T}P_u(\Lambda))\Pi(\Lambda)\\&+\Pi(\Lambda)(A+B_u\Lambda{+}B_aR^{-1}B_a^\mathrm{T}P_u(\Lambda))^\mathrm{T}{-}\bar{K}^\mathrm{T}R^{-1}B_a^\mathrm{T}\label{eq:grad_Pi}\\&-B_aR^{-1}\bar{K}+P_u(\Lambda)B_aR^{-2}B_a^\mathrm{T}+B_aR^{-2}B_a^\mathrm{T}P_u(\Lambda)=0.\nonumber
\end{align}
\end{lemma}
 According to the notation and results of Lemma \ref{le:grad}, from lines 9-10 of Algorithm \ref{al:BSOR}, we have
\begin{align*}
\omega(\Lambda^\mathrm{GS}-\Lambda^i)&=\omega(-\Gamma^{-1}B_u^\mathrm{T}P_u^{i+1}\Pi^{i+1}-\Lambda^i)\\&=\omega(-\Gamma^{-1}B_u^\mathrm{T}P_u(\Lambda^i)\Pi(\Lambda^i)-\Lambda^i)\\&=-\frac{1}{2}\omega\Gamma^{-1}\cdot2(B_u^\mathrm{T}P_u(\Lambda^i)\Pi(\Lambda^i)+\Gamma\Lambda^i)\\&=-\frac{1}{2}\omega\Gamma^{-1}\frac{\mathrm{d}\tilde{J}(\Lambda^i)}{\mathrm{d}\Lambda^i}.
\end{align*}
Hence, we can interpret the iterative update of the deception gain $\Lambda^i$ in line 10 of Algorithm \ref{al:BSOR} as
\begin{equation}\label{eq:grad_update}
\Lambda^{i+1}\leftarrow \Lambda^{i}-\frac{1}{2}\omega\Gamma^{-1}\frac{\mathrm{d}\tilde{J}(\Lambda^i)}{\mathrm{d}\Lambda^i},
\end{equation}
where we recall that $\tilde{J}(\Lambda)$ is the cost function in \eqref{eq:opt1} when $P_u$ is viewed implicitly as a function of $\Lambda$ (see \eqref{eq:JL}). This means that Algorithm \ref{al:BSOR} is a first-order method on the ``generalized" cost \eqref{eq:JL} and should, in principle, converge to a stationary point, which would then correspond to a stationary point of problem \eqref{eq:opt1}. However, a couple of issues must be considered.

The first issue is that the domain of  $\tilde{J}$ is not equal to all of $\mathbb{R}^{m_u\times n}$. This is because $\Lambda$ must be such that a solution $P_u(\Lambda)\in\mathbb{S}_+$ for \eqref{eq:PL} exists, and this is clearly not the case for all $\Lambda$ in $\mathbb{R}^{m_u\times n}$. Accordingly, if Algorithm \ref{al:BSOR} happens to abruptly exit the domain of $\tilde{J}$ at some iteration, it will terminate prematurely with an error while attempting to solve \eqref{eq:it1}. The second issue is that even if the iterates $\Lambda^i$ do not violate the domain of $\tilde{J}$ throughout the execution of Algorithm \ref{al:BSOR}, convergence to a stationary point can be assured only as long as the gradient of $\tilde{J}$ has a bounded Lipschitz constant \cite{bertsekas1997nonlinear}. Fortunately, in what follows,
we will show that neither of those two issues will become an obstacle to convergence, under certain conditions.

Towards this end, let us define the set
\begin{equation}\label{eq:S0}
\mathcal{S}_0=\{\Lambda\in\mathbb{R}^{m_u\times n}~|~\exists P_{u}(\Lambda)\in\mathbb{S}_+~\textrm{and}~\tilde{J}(\Lambda)\le \tilde{J}(0)\},
\end{equation}
where we observe that $\Lambda^0$, i.e., the starting point of Algorithm \ref{al:BSOR}, lies inside $\mathcal{S}_0$.  The following auxiliary lemma shows that both $P_{u}(\Lambda)$ and $\Pi(\Lambda)$ are bounded uniformly in $\mathcal{S}_0$, and that $A+B_u\Lambda+B_aK_u(\Lambda)$ remains strictly stable uniformly in $\mathcal{S}_0$.

\begin{lemma}\label{le:uniformly_bounded}
Let the following strengthened version of \eqref{eq:stab_condition} hold:
\begin{equation}\label{eq:stab_condition2}
\norm{K^\star-\bar{K}}_{\mathrm{F}}\le \frac{\epsilon}{2}\frac{\sqrt{\lambda_{\textrm{min}}(\Gamma)}}{\sqrt{\lambda_{\textrm{min}}(\Gamma)}\norm{SB_a}_{\mathrm{F}}+\norm{SB_u}_{\mathrm{F}}},
\end{equation}
where $S$ is as in \eqref{eq:LE_S}, and $\epsilon\in(0,1)$.
Then, the following hold.
\begin{enumerate}
\item $A+B_u\Lambda+B_aK_u(\Lambda)$ is strictly stable uniformly in $\mathcal{S}_0$.  Specifically, $\sup_{\Lambda\in\mathcal{S}_0}\alpha(A+B_u\Lambda+B_aK_u(\Lambda))<0$, and, more strongly, 
\begin{multline}\label{eq:uni_stable}
(A+B_u\Lambda+B_aK_u(\Lambda))^\mathrm{T}S\\+S(A+B_u\Lambda+B_aK_u(\Lambda))\preceq -2(1-\epsilon) I
\end{multline} 
for all $\Lambda\in \mathcal{S}_0$, with $S$ as in \eqref{eq:LE_S}.
\item There exist finite constants $\bar{P},~\bar{\Pi}>0$ such that $\norm{P_u(\Lambda)}_\mathrm{F}\le \bar{P}$ and $\norm{\Pi(\Lambda)}_\mathrm{F}\le \bar{\Pi}$ for all $\Lambda\in\mathcal{S}_0$.
\end{enumerate}
\end{lemma}

The practical interpretation of Lemma \ref{le:uniformly_bounded} is that, by continuity,  $P_u(\Lambda)\in\mathbb{S}_+$ exists not only everywhere in $\mathcal{S}_0$ but also in a neighborhood around it. Therefore, the negated gradient $-\frac{\mathrm{d}\tilde{J}(\Lambda)}{\mathrm{d}\Lambda}$ points in a direction towards which $P_u(\Lambda)\in\mathbb{S}_+$ exists, and hence this direction is in the interior of $\mathcal{S}_0$. Accordingly, Algorithm \ref{al:BSOR} will always remain within a set where both equations \eqref{eq:it1} and \eqref{eq:it2} admit unique solutions and will not terminate prematurely. This effectively deals with one of the two issues raised in the preceding paragraphs.

Next, we focus on the second issue raised in the preceding discussion; that the gradient of $\tilde{J}$ must admit a global Lipschitz constant on $\mathcal{S}_0$ for Algorithm \ref{al:BSOR} to converge to a stationary point of \eqref{eq:opt1}. Generally speaking, such a conclusion cannot be directly obtained by inspection because $P_u(\Lambda)$ and $\Pi(\Lambda)$ in \eqref{eq:grad_formula} are derived as solutions of nonlinear equations, and hence can depend sublinearly on $\Lambda$. Nevertheless, using the uniform boundedness result asserted in the second part of Lemma \ref{le:uniformly_bounded}, we are able to establish such property as follows.

\begin{lemma}\label{le:lipschitz}
Let \eqref{eq:stab_condition2} hold. Then, the gradient of the cost function $\tilde{J}(\Lambda)$, i.e., $\frac{\mathrm{d}\tilde{J}(\Lambda)}{\mathrm{d}\Lambda}$, is globally Lipschitz on $\Lambda\in\mathcal{S}_0$. 
\end{lemma}

Finally, we combine the previous Lemmas to obtain the main result of this subsection, which characterizes the convergence properties of Algorithm \ref{al:BSOR}. In particular, we show that Algorithm 1 converges to a stationary point of \eqref{eq:opt1}, and that this point is a minimum under a stricter condition on the regulation matrix $\Gamma$.
\begin{theorem}\label{th:convergence}
Let \eqref{eq:stab_condition2} hold and $\omega<\frac{2}{L}\frac{\lambda_{\mathrm{min}}(\Gamma^{-1})}{\lambda_{\mathrm{max}}^2(\Gamma^{-1})}$, where $L$ is the Lipschitz constant of $\frac{\mathrm{d}\tilde{J}(\Lambda)}{\mathrm{d}\Lambda}$ on $\mathcal{S}_0$. Consider the sequence of matrices $\{P_u^i,\Pi^i,\Lambda^i\}_{i\in\mathbb{N}}$ generated by Algorithm \ref{al:BSOR}. Then:
\begin{enumerate}
\item At each step $i\in\mathbb{N}$, there exist unique solutions $P_u^{\mathrm{GS}}\in\mathbb{S}_+,~\Pi^{\mathrm{GS}},~\Lambda^{\mathrm{GS}}$ to the equations \eqref{eq:it1}-\eqref{eq:it3};
\item $\tilde{J}(\Lambda^{i+1})\le \tilde{J}(\Lambda^{i})$ for all $i\in\mathbb{N}$;
\item $\lim_{i\rightarrow\infty}\frac{\mathrm{d}\tilde{J}(\Lambda^i)}{\mathrm{d}\Lambda}=0$, that is, the stationarity equations \eqref{eq:eq1}-\eqref{eq:eq3} hold in the limit, and Algorithm \ref{al:BSOR} terminates for any tolerance $\epsilon>0$;
\item There exists $\gamma^\star>0$, such that if $\Gamma\succ\gamma^\star I$ then Algorithm \ref{al:BSOR} converges to a minimum of \eqref{eq:opt1}. 
\end{enumerate}
\end{theorem}

\begin{remark}
An alternative to having a small constant step size $\omega$ in Algorithm \ref{al:BSOR}, would be to instead have an iteration-varying step $\omega^i$ with the property that $\sum_{i\in\mathbb{N}}(\omega^{i})^2<\infty$  and $\sum_{i\in\mathbb{N}}\omega^i=\infty$. While such a choice retains convergence guarantees, empirical results indicate that it makes Algorithm $1$ converge substantially more slowly. 
\end{remark}

\section{Discussion}\label{sec:discuss}

In this section, we discuss how we can extend the results of the preceding sections to tackle more general cases of deception in linear-quadratic control.

\subsection{On the Condition of Lemma \ref{le:stab}}\label{sec:discuss1}

A sufficient condition for guaranteeing the existence of a minimizer to \eqref{eq:opt1} is inequality \eqref{eq:stab_condition}, requiring that the attack gain $\bar{K}$ the defender wants to induce be sufficiently close to the original gain $K^\star$. As stated in the preceding sections, such a condition can be conservative because we derived it without making any a priori assumptions regarding the deceptive target gain $\bar{K}$. 
On the other hand, if we consider the possibility that $\bar{K}$ will be better for the defender in terms of closed-loop performance when compared to the original gain $K^\star$, then \eqref{eq:opt1} is unnecessary. We illustrate this in the following example, wherein the solution to the ARE \eqref{eq:are} that yields the nominal attack $K^\star$ does not even exist, because $\lambda_{\textrm{min}}(R)$ is small and hence $K^\star$ is destabilizing.

\begin{example}
Consider a scalar system where the closed-loop state matrix is $A=-1$, the input matrices are $B_u=B_a=1$, and the state cost matrix is $Q=1$. Then, the ARE \eqref{eq:are} reduces to the scalar quadratic equation
\begin{equation}\label{eq:are_example}
P^2-2RP+R=0.
\end{equation}
When $R>1$, the solutions to this equation are given by $P_{1,2}=R\pm\sqrt{R-1}\sqrt{R}$, and hence the minimal positive definite solution is $P=R-\sqrt{R-1}\sqrt{R}$. Therefore, from \eqref{eq:K}, the optimal attack gain is $K^\star=1-\sqrt{1-R^{-1}}$. On the other hand, when $0<R< 1$, \eqref{eq:are_example} admits no solutions, the value of \eqref{eq:Jnom} is unbounded, and the optimal attack $K^\star$ destabilizing. Yet, the defender can deceive the attacker into a milder attack that renders the closed-loop stable, while bypassing the rationale of Lemma \ref{le:stab}.

More specifically, let $R=0.5<1$ and suppose that the target attack gain the defender wants to induce is $\bar{K}=0.2$; a milder target than $K^\star$ which does not disrupt closed-loop stability. Then, assuming $\Gamma=0$ and thus further violating \eqref{eq:stab_condition}, the defender would need to find the gain $\Lambda$ such that $K_u(\Lambda)=R^{-1}P_u(\Lambda)=\bar{K}=0.2$, and thus $P_u(\Lambda)=0.1$. Plugging this in \eqref{eq:RicL} yields  $\Lambda=-4.1$. 
Therefore, $A+B_u\Lambda+B_aK_u(\Lambda)=A+B_u\Lambda+B_a\bar{K}=-4.9$, i.e., the minimum of \eqref{eq:opt1} is attained and \eqref{eq:RicL} admits a stabilizing solution. In addition, $A+B_aK_u(\Lambda)=-1+0.2=-0.8$, i.e., the deceptive attack gain does not disrupt closed-loop stability. 
\end{example}

The example above showcases that even when \eqref{eq:stab_condition} does not hold, \eqref{eq:opt1} and its ARE constraint \eqref{eq:RicL} can still be well-defined. The main idea is that it is not that the target $\bar{K}$ needs to be close to $K^\star$ per se, as \eqref{eq:stab_condition} requires; rather, that it needs to be equal to an attack gain that is milder than $K^\star$, or at least not much worse than $K^\star$. Accordingly, in these cases, one should select the initialization $\Lambda^0$ in Algorithm $1$  to render $A+B_u\Lambda^0$ sufficiently stable, i.e., with a sufficiently small spectral abscissa\footnote{This implicitly requires that $(A,B_u)$ be controllable.}. In this way, one can guarantee that the first ARE iterate \eqref{eq:it1} will admit a well-defined stabilizing solution, and that \eqref{eq:it2} will admit a unique solution, too.

We capture the intuition that \eqref{eq:stab_condition} does not need to hold when $\bar{K}$ is milder than $K^\star$ in the technical result below, for the case $\bar{K} = 0$.
\begin{proposition}
Let $\bar{K}=0$ and suppose $\mathrm{Ran}(B_a)\subseteq\mathrm{Ran}(B_u)$. Then, there exists $\gamma^\star>0$ such that if $\Gamma\prec \gamma^\star I$ then \eqref{eq:opt1} admits a global minimizer.
\end{proposition}

\begin{remark}
The condition $\mathrm{Ran}(B_a)\subseteq\mathrm{Ran}(B_u)$ requires that the adversary does not have more channels to attack the system than the defender has to control it. 
\end{remark}

\subsection{On the Assumption of Known $Q$ and $R$ Matrices}\label{sec:discuss2}

In Assumption \ref{ass:2}, we assumed the defender does not know the learning algorithm used by the adversary, but knows its objective, i.e., its matrices $Q$ and $R$. A natural question thus regards how well the deception scheme will work when the actual adversary matrices are different.

To investigate this question, suppose that the actual weighting matrices in the adversary's objective are equal to $\hat{Q}\ne Q$ and $\hat{R}\ne R$. In addition,
consider the cost function \eqref{eq:opt1}, but under these actual weighting matrices $\hat{Q},~\hat{R}$ of the adversary:
\begin{align}\label{eq:hatL}
&\hat{J}(\Lambda)=\norm{\hat{K}_u(\Lambda)-\bar{K}}_{\mathrm{F}}^2+\mathrm{tr}
(\Lambda^\mathrm{T}\Gamma\Lambda),
\end{align}
where $\hat{K}_u(\Lambda)=\hat{R}^{-1}B_a^\mathrm{T}\hat{P}_u(\Lambda)$ and
\begin{multline}\label{eq:Phat}
    (A+B_u\Lambda)^\mathrm{T}\hat{P}_u(\Lambda)+\hat{P}_u(\Lambda)(A+B_u\Lambda)+\hat{Q}\\+\hat{P}_u(\Lambda)B_a\hat{R}^{-1}B_a^\mathrm{T}\hat{P}_u(\Lambda)=0, ~\hat{P}_u(\Lambda)\in\mathbb{S}_+.
\end{multline}
In the following proposition, we show that when $\bar{K}=0$, the gain the defender derives by optimizing \eqref{eq:opt1} is still efficient in deceiving the adversary towards $\bar{K}=0$, despite the incorrectly assumed matrices $Q$ and $R$ in \eqref{eq:opt1}. In particular, we show that i) in the case the actual matrices of the adversary satisfy $\hat{Q}\prec Q$ and $\hat{R}\succ R$, then the optimal value of \eqref{eq:opt1} provides an upper bound for \eqref{eq:hatL}; and ii) in the general case, the optimal value of \eqref{eq:opt1} provides an approximate upper bound for \eqref{eq:hatL}. 

\begin{proposition}\label{pr:2}
 Let $\bar{K}=0$. Denote as $\Lambda^\star$ the minimizer of \eqref{eq:opt1}. The following hold true.
\begin{enumerate}
\item If $\hat{Q}\preceq Q$ and $\hat{R}\succeq R$ then $\hat{J}(\Lambda^\star)\le \tilde{J}(\Lambda^\star)$.
\item  Suppose $\mathrm{Ran}(B_a)\subseteq\mathrm{Ran}(B_u)$. Then, for every $\epsilon>0$ there exists $\gamma^\star>0$ such that if $\Gamma\preceq\gamma^\star I$ then $\hat{J}(\Lambda^\star)\le \tilde{J}(\Lambda^\star)+\epsilon$.
\end{enumerate}
\end{proposition}

\begin{remark}
Proposition~2 establishes that $\hat{J}(\Lambda^\star) \leq \tilde{J}(\Lambda^\star) + \epsilon$. 
The gap $\hat{J}(\Lambda^\star) - \tilde{J}(\Lambda^\star)$ depends continuously on the cost-matrix mismatches $\|\hat{Q} - Q\|_\textrm{F}$ and $\|\hat{R} - R\|_\textrm{F}$ and vanishes as these mismatches vanish. This follows from standard norm inequalities together with the continuous dependence of the stabilizing algebraic Riccati solution (and hence of the induced feedback gain) on its cost matrices.
\end{remark}

\subsection{On the Deception against Minimizing Data-Driven Linear-Quadratic Regulators}\label{sec:minimizing}

A dual setup in data-driven linear quadratic control is one where a defender wants to learn the solution to the minimization problem
\begin{align}\nonumber
     \min_{N\in\mathbb{R}^{m_u\times n}}  ~&\int_0^\infty(x^\mathrm{T}(\tau)Qx(\tau)+u^\mathrm{T}(\tau)Mu(\tau))\mathrm{d}\tau,\\ 
     \textrm{s.t.} \qquad &\dot{x}(t)=Ax(t)+B_uu(t),\label{eq:Jper_u}\\
     &u(t)=Nx(t),\nonumber
\end{align}
with $Q,~M\succ0$.
Assuming $(A,B_u)$ is stabilizable, then we know the solution to this problem is
\begin{equation}\label{eq:K2}
N^\star=-M^{-1}B_u^\mathrm{T}Z,
\end{equation}
where $Z\succ0$ is the positive definite solution of the ARE
\begin{equation}\label{eq:are_u}
A^\mathrm{T}Z+ZA+Q-ZB_uM^{-1}B_u^\mathrm{T}Z=0.
\end{equation}
Accordingly, a dual setup in deceiving such a defender would be to inject an adversarial perturbation $a(t)=Lx(t)$ in the system, forcing the defender to converge to the incorrect  linear-quadratic gain
\begin{equation}\label{eq:KL2}
N_a(L)=-M^{-1}B_u^\mathrm{T}Z_a(L),
\end{equation}
where $Z_a(L)\succ0$ is obtained by solving the  perturbed  ARE
\begin{multline}\label{eq:are_u2}
(A+B_aL)^\mathrm{T}Z_a+Z_a(A+B_aL)\\+Q-Z_aB_uM^{-1}B_u^\mathrm{T}Z_a=0.
\end{multline}
Note that \eqref{eq:are_u} is much easier to analyze than \eqref{eq:RicL}; as long as $\mathrm{Ran}(B_a)\subseteq\mathrm{Ran}(B_u)$, i.e., as long as the attacker does not have more input channels than the defender, then the positive definite solution to  \eqref{eq:are_u2} always exists (see \cite{fotiadis2024poisoning}). Therefore, it is straightforward to verify that the dual deception problem
\begin{align}\nonumber
&\inf_{L\in\mathbb{R}^{m_a\times n},Z_a\in\mathbb{R}^{n\times n}}\norm{N_a(L)-\bar{N}}_{\mathrm{F}}^2+\mathrm{tr}(L^\mathrm{T}\Gamma L),\\
&~\quad\textrm{s.t.} \qquad\quad \eqref{eq:KL2}, \eqref{eq:are_u2}, ~ Z_a\succ0, \label{eq:opt2}
\end{align}
admits a global minimizer for any choice of $\Gamma\succ0$ and $\bar{N}\in\mathbb{R}^{m_u\times n}$. With minor modifications, one may then adjust Algorithm \ref{al:BSOR} to solve \eqref{eq:opt2} numerically (see \cite{fotiadis2024poisoning}). The guarantees of convergence will follow the same line as those of Section \ref{sec:conv}, but will not be constrained by the condition of Lemma \ref{le:stab} since \eqref{eq:opt2} is well-defined for any $L\in\mathbb{R}^{m_a\times n}$.

\section{Simulations}\label{sec:sim}

We consider a linearized model of the ADMIRE
benchmark aircraft,  whose open-loop state and input matrices $A_o, B_u$ are given in \cite{admire}.
Using these, we define the closed-loop state matrix of the aircraft as $A=A_o+B_uF$, where the gain $F$ originates
from a linear-quadratic regulation problem with identity weighting matrices.

\begin{figure}[!t] 
		\centering
                \includegraphics[width=8.3cm,height=4.5cm]{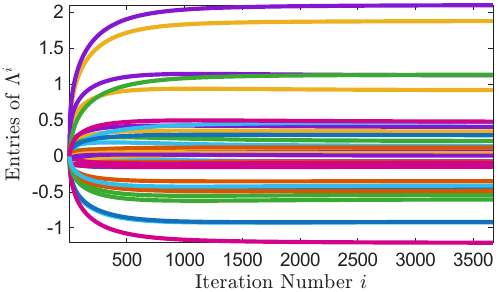}
                \caption{\small Case 1: The evolution of the entries of $\Lambda^i$ during Algorithm \ref{al:BSOR}.}
                \label{fig:L}
\end{figure}
\begin{figure}[!t] 
		\centering
                \includegraphics[width=8cm,height=4.5cm]{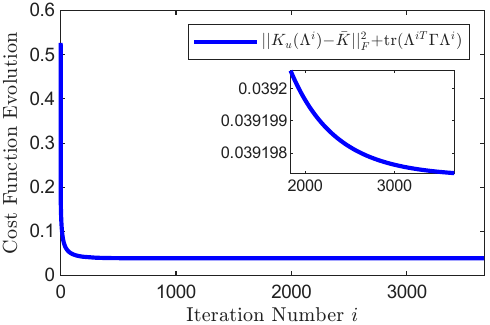}
                \caption{\small Case 1: The evolution of the cost \eqref{eq:opt1} during  Algorithm \ref{al:BSOR}.}
                \label{fig:cost}
\end{figure}

\begin{figure}[!t] 
		\centering
                \includegraphics[width=8.5cm,height=4.5cm]{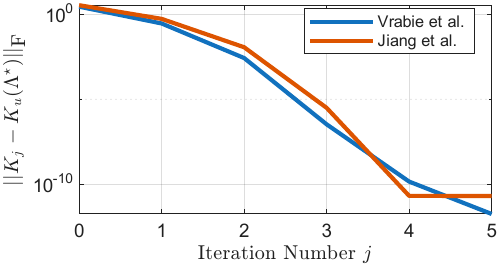}
                \caption{\small Impact of deception on the evolution of the learning algorithms presented in \cite{vrabie2009adaptive, jiang2012computational}. The figure shows the distance of the learnt gain $K_j$ at each iteration $j$ of the learning algorithm, from the suboptimal gain $K_u(\Lambda^\star)$.  }
                \label{fig:learning}
\end{figure}

\subsection{Case 1: Deception against a Performance-Degrading Attack}

In the first case, we consider that an adversary has access to the first and the third actuator of the aircraft. Since each actuator corresponds to a column of $B_u$, this means that 
\begin{equation*}\small
    B_a^\textrm{T}=\begin{bmatrix}-0.0062 & -0.0072 & 1.2456 & 2.7172 & -0.7497 \\
    -0.0709 & 0.0039 & -10.6058 & -2.4724 & -0.4923 \\\end{bmatrix}.
\end{equation*}
The adversary then wants to use these actuators to disturb the stabilization of the aircraft. To that end, it seeks to learn the attack that solves the optimization problem \eqref{eq:Jnom} with $Q=I_5$ and $R=0.6I_2$.

On the other side, the defender who is aware of the adversary's objective wants to deceive them to converge as closely as possible to the attack gain $\bar{K}=0$. To that end, the defender plugs a deceptive gain in the system to corrupt the attacker's learning data. This gain is obtained by performing Algorithm $1$ with $\omega=0.001$, so as to numerically solve the deception optimization \eqref{eq:opt1} with $\Gamma=10^{-3}I_7$.

The evolutions of the deception gain sequence $\Lambda^i$ generated by Algorithm \ref{al:BSOR}, as well as of the deception cost \eqref{eq:opt1}, are shown in Figures \ref{fig:L}-\ref{fig:cost}. From Figure \ref{fig:L}, we notice that Algorithm \ref{al:BSOR} reaches a stationary point of the deception problem \eqref{eq:opt1} after about 3500 iterations, which validates Theorem \ref{th:convergence}. In addition, Figure \ref{fig:cost} shows that the deception cost \eqref{eq:opt1} is decreasing at each iteration of Algorithm \ref{al:BSOR}. This is in line with the observation made in Lemma \ref{le:grad}, regarding the Algorithm's equivalence to a first-order gradient method. It indicates that, at each iteration $i$, the gain $\Lambda^i$ the defender computes becomes more and more efficient in deceiving the adversary.

\begin{table}[!t]
  \begin{center}
  \caption{\small Ratio $[K_u(\Lambda^\star)]_{ij}/[K^\star]_{ij}$ between optimal and deceived attack gains. \label{tab:ratio}}
  \resizebox{\columnwidth}{!}{%
   \begin{tabular}{c||c|c|c|c|c}
      \toprule % <-- Toprule here
      \textbf{ } & Col. $1$ & Col. $2$ & Col. $3$ & Col. $4$ & Col. $5$\\
      \midrule % <-- Midrule here
      Row $1$  & $0.0827$ & $0.9429$ & $0.0545$ & $0.2748$ & $0.2439$\\
      Row $2$ & $0.1581$ & $0.0046$ & $0.2179$ & $0.1103$ & $0.2189$ \\
      \bottomrule % <-- Bottomrule here
    \end{tabular}}
  \end{center}
\end{table}

\begin{table}[!t]
  \begin{center}
  \caption{\small Ratio $[\hat{K}_u(\Lambda^\star)]_{ij}/[\hat{K}^\star]_{ij}$ between optimal and deceived attack gains, under incorrectly assumed weighting matrices for the adversary. \label{tab:ratio2}}
  \resizebox{\columnwidth}{!}{%
   \begin{tabular}{c||c|c|c|c|c}
      \toprule % <-- Toprule here
      \textbf{ } & Col. $1$ & Col. $2$ & Col. $3$ & Col. $4$ & Col. $5$\\
      \midrule % <-- Midrule here
      Row $1$  & $0.0655$ & $-0.0829$ & $0.0355$ & $0.2159$ & $0.2531$\\
      Row $2$ & $0.1190$ & $0.0014$ & $0.1761$ & $0.0694$ & $0.1366$  \\
      \bottomrule % <-- Bottomrule here
    \end{tabular}}
  \end{center}
\end{table}

\begin{table}[!t]
  \centering
  \caption{\small  State energy 
  $\mathcal{E}_x = \int_{0}^{\infty} \norm{x(t)}^2\, dt$ 
  under nominal and deceived attacks for different adversarial input weights $R$. 
  The value $\infty$ indicates closed-loop instability. }
  \label{tab:energy}
  \resizebox{\columnwidth}{!}{%
  \begin{tabular}{c||c|c|c|c}
    \toprule
    \textbf{Case} 
    & $R=0.2 I_2$ 
    & $R=0.4 I_2$ 
    & $R=0.6 I_2$ 
    & $R=0.8 I_2$ \\
    \midrule
    Nominal
    & $\infty$ 
    & $0.022844$ 
    & $0.008229$ 
    & $0.008055$ \\
    Deceived 
    & $0.007948$ 
    & $0.007924$ 
    & $0.007914$ 
    & $0.007909$ \\
    \bottomrule
  \end{tabular}}
\end{table}

Table \ref{tab:ratio} additionally shows the ratio between the optimal attack $K^\star$ the attacker would have learned in the absence of deception, and the attack $K_u(\Lambda^\star)$ they were forced to learn instead. Clearly, the latter has been suppressed substantially and is thus more benign in disturbing closed-loop performance. Note that $K_u(\Lambda^\star)$ does not become exactly zero because of the regularization term that penalizes large values of $\Lambda$ in \eqref{eq:opt1}.

To validate this prediction at the level of the attacker’s learning dynamics, Figure \ref{fig:learning} illustrates how the deception gain $\Lambda^\star$ affects the evolution of two learning algorithms from the literature \cite{vrabie2009adaptive, jiang2012computational}. One assumes unknown $A$ and $B_a$ matrices \cite{jiang2012computational}, whereas the other assumes unknown $A$ and known $B_a$ \cite{vrabie2009adaptive}. In both cases, we observe the learning process converges to the suboptimal attack $K_u(\Lambda^\star)$ that the defender had predicted.

To further assess the impact at the level of closed-loop performance, 
Table \ref{tab:energy} reports the resulting state energy 
$\mathcal{E}_x = \int_{0}^{\infty} \norm{x(t)}^2 dt$ 
under nominal and deceived attacks for different values of $R$. 
We observe that, for small values of $R$, the nominal attack may even destabilize the system (see next subsection), 
whereas the deceived attack consistently yields finite and smaller energy.

\begin{figure}[!t]
\centering
\includegraphics[width=0.44\textwidth]{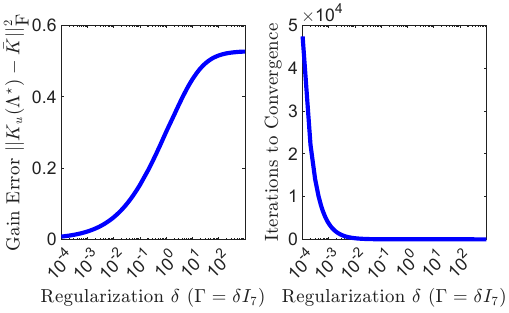}
\caption{\small Effect of the regularization parameter $\Gamma=\delta I_7$ on the gain error $\|K_u(\Lambda^\star)-\bar{K}\|_F^2$ (left) and the number of iterations required by Algorithm 1 to converge (right).}
\label{fig:gaindif}
\end{figure}

To further investigate the role of the regularization parameter $\Gamma=\delta I_7$, we evaluate the resulting gain error $\norm{K_u(\Lambda^\star)-\bar K}_\mathrm{F}^2$ for different values of $\delta$, together with the number of iterations required by Algorithm 1 to converge. The results are shown in Figure \ref{fig:gaindif}. We observe that smaller values of $\delta$ allow the defender to deceive the adversary more closely toward the target attack gain $\bar K=0$, resulting in lower gain error. Conversely, increasing $\delta$ reduces deception effectiveness but improves the numerical convergence of Algorithm \ref{al:BSOR}. This illustrates the practical trade-off between achieving low deception error and convexification.

Finally, we showcase the robustness of the derived deception gain when the adversary's weighting matrices are different than those assumed. Specifically, we take the deception gain we previously derived under $Q=I_5$ and $R=0.6I_2$, and show that even if the adversary's weighting matrices are equal to $\hat{Q}=5I_5\ne Q$ and $\hat{R}=2I_2\ne R$, then the learned adversarial attack is still suppressed. We show the ratio between the learned adversarial attack $\hat{K}_u(\Lambda^\star)$ and the nominal attack $\hat{K}^\star$ under $\hat{Q},~\hat{R}$ in Table \ref{tab:ratio2}, validating this conjecture along with the analysis we carried out in Section \ref{sec:discuss2}.

\subsection{Case 2: Deception against a Destabilizing Attack}

In Case 2, we consider the same setup for the adversary, but where $R=0.1I_2$ in their optimization problem \eqref{eq:Jnom}. In this case, the optimal attack $K^\star$ is destabilizing. To find a deception gain that will force the adversary to learn a more benign attack instead, we re-employ the same scheme as in Case 1 for the defender. However, following the discussion in Section \ref{sec:discuss}, we select $\Lambda^0$ in Algorithm $1$ so that it makes the eigenvalues of $A+B_u\Lambda^0$ sufficiently negative; in particular, equal to $-100$.

\begin{figure}[!t] 
		\centering
                \includegraphics[width=8cm,height=4.8cm]{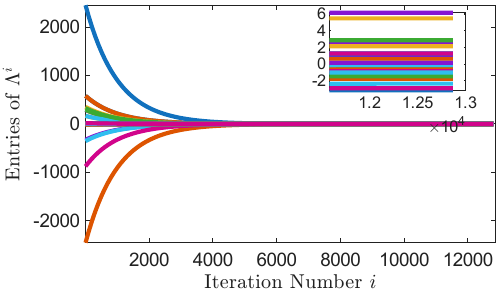}
                \caption{\small Case 2: The evolution of the entries of $\Lambda^i$ during Algorithm \ref{al:BSOR}.}
                \label{fig:L2}
\end{figure}
\begin{figure}[!t] 
		\centering
                \includegraphics[width=8cm,height=4.8cm]{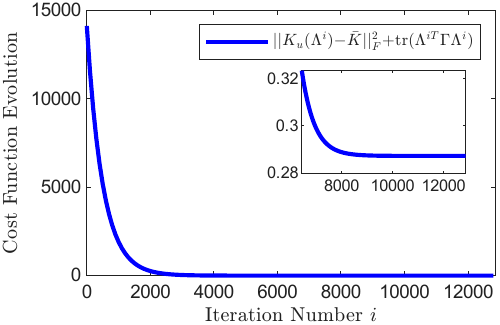}
                \caption{\small Case 2: The evolution of the cost \eqref{eq:opt1} during  Algorithm \ref{al:BSOR}.}
                \label{fig:cost2}
\end{figure}

The evolution of the deception gain sequence $\Lambda^i$ and the deception cost \eqref{eq:opt1} during Algorithm \ref{al:BSOR} is shown in Figures \ref{fig:L2}-\ref{fig:cost2}. While we still observe a decreasing deception cost and a convergent deception gain, the number of iterations needed to achieve the same level of tolerance increased significantly. This was due to the conservative initialization of $\Lambda^0$ very far from the optimal deception gain $\Lambda^\star$. Meanwhile, the eigenvalues of the state matrix in closed-loop with the deceptive attack gain are equal to $-17.6911, -7.7253, -2.5628 + 0.4642i, -2.5628 - 0.4642i, -1.6568$. This means that the learned optimal attack under deception is not destabilizing, as opposed to the nominal optimal attack $K^\star$.

\section{Conclusion}\label{sec:conc}

We study the problem of deception in the context of data-driven linear-quadratic control. Given an adversary (defender) who wants to learn the optimal linear-quadratic attack (control) for a system using data, we design a deceptive input that optimally corrupts this data and forces the learner to converge to a prespecified suboptimal policy. We characterize the deceptive input as a solution of coupled matrix equations, which we subsequently solve using fixed-point iteration and with convergence guarantees. 

Future work includes an extension of the presented setup to the deception of nonlinear data-driven optimal control.

\bibliographystyle{ieeetr}      
\bibliography{references.bib}

\appendix

The following lemma presents matrix differentiation formulas from \cite{bernstein2009matrix}. Note that since \cite{bernstein2009matrix} follows the convention $\left[\frac{\mathrm{d}}{\mathrm{d}X}\right]_{ij}=\frac{\mathrm{d}}{\mathrm{d}[X]_{ji}}$, these formulas have been transposed.
\begin{lemma}
The following formulas hold true by \cite{bernstein2009matrix}
\begin{itemize}
    \item For any $A\in\mathbb{R}^{n\times m}, B\in \mathbb{R}^{l\times n}$:
\begin{align}
&X\in\mathbb{R}^{m\times n}:~\frac{\mathrm{d}}{\mathrm{d}X}\mathrm{tr}AX=\frac{\mathrm{d}}{\mathrm{d}X}\mathrm{tr}XA=A^\mathrm{T},\label{eq:AX}\\
&X\in\mathbb{R}^{m\times l}:~\frac{\mathrm{d}}{\mathrm{d}X}\mathrm{tr}AXB=A^\mathrm{T}B^\mathrm{T}, \label{eq:AXB}\\
&X\in\mathbb{R}^{l\times m}:~\frac{\mathrm{d}}{\mathrm{d}X}\mathrm{tr}AX^\mathrm{T}B=BA.\label{eq:AXtB}
\end{align}
\item For any $A\in\mathbb{R}^{n\times m}, B\in \mathbb{R}^{m\times n},  X\in\mathbb{R}^{m\times m}$:
\begin{equation}
   \frac{\mathrm{d}}{\mathrm{d}X}\mathrm{tr}AX^2B=
   A^\mathrm{T}B^\mathrm{T}X^\mathrm{T}+X^\mathrm{T}A^\mathrm{T}B^\mathrm{T}.\label{eq:AX2B} \\
\end{equation}
\item For any  $A,B\in\mathbb{R}^{n\times m}, X\in\mathbb{R}^{m\times n}$:  
\begin{equation}
    \frac{\mathrm{d}}{\mathrm{d}X}\mathrm{tr}AXBX=A^\mathrm{T}X^\mathrm{T}B^\mathrm{T}+B^\mathrm{T}X^\mathrm{T}A^\mathrm{T}.\label{eq:AXBX}
\end{equation}
\item For any $A\in\mathbb{R}^{n\times n}, B\in \mathbb{R}^{m\times m}, X\in\mathbb{R}^{n\times m}$:
\begin{equation}
    \frac{\mathrm{d}}{\mathrm{d}X}\mathrm{tr}AXBX^\mathrm{T}=AXB+A^\mathrm{T}XB^\mathrm{T}.\label{eq:AXBXt}
\end{equation}
\end{itemize}
\end{lemma}

The proofs of the main results follow next.

\textit{Proof of Lemma \ref{le:stab}}:
    Let $\{\hat{\Lambda}^n,\hat{P}_u^n\}_{n\in\mathbb{N}}$ be a minimizing sequence of \eqref{eq:opt1}, i.e., a sequence that satisfies the constraints of \eqref{eq:opt1}, and for which $J(\hat{\Lambda}^n,\hat{P}_u^n)$ converges to the infimum of \eqref{eq:opt1} as $n\rightarrow\infty$. Note that $\hat{\Lambda}^n$ and $K_u(\hat{\Lambda}^n)$ must be bounded uniformly; for if they are not, then $\lim_{n\rightarrow\infty}J(\hat{\Lambda}^n,\hat{P}_u^n)=\infty$ owing to the fact that $J(\hat{\Lambda}^n,\hat{P}_u^n)\ge\norm{K_u(\hat{\Lambda}^n)-\bar{K}}_{\mathrm{F}}^2+\lambda_{\textrm{min}}(\Gamma)\norm{\hat{\Lambda}^n}^2_{\mathrm{F}}$, whereas $J(0,P)<\infty$, which contradicts optimality. Therefore, one can extract from  $\{\hat{\Lambda}^n,\hat{P}_u^n\}_{n\in\mathbb{N}}$ a subsequence  $\{\bar{\Lambda}^n,\bar{P}_u^n\}_{n\in\mathbb{N}}$ such that $\lim_{n\rightarrow\infty}\bar{\Lambda}^n=\bar{\Lambda}\in\mathbb{R}^{m_u\times n}$ and $\lim_{n\rightarrow\infty}K_u(\bar{\Lambda}^n)=\bar{K}_u\in\mathbb{R}^{m_a\times n}$ [Theorem 3.6 in \cite{rudin1964principles}].  Using these limits and \eqref{eq:LE_S}, we have
    \begin{align*}       (A&+B_u\bar{\Lambda}+B_a\bar{K}_u)^\mathrm{T}S+S(A+B_u\bar{\Lambda}+B_a\bar{K}_u)\\=&(A+B_aK^\star)^\mathrm{T}S+S(A+B_aK^\star)+SB_u\bar{\Lambda} +(SB_u\bar{\Lambda} )^\mathrm{T}\\&+SB_a(\bar{K}_u-K^\star)+(SB_a(\bar{K}_u-K^\star))^\mathrm{T}\\=&-2I+SB_u\bar{\Lambda} +(SB_u\bar{\Lambda} )^\mathrm{T}\\&+SB_a(\bar{K}_u-K^\star)+(SB_a(\bar{K}_u-K^\star))^\mathrm{T}.
    \end{align*}
    Based on the Lyapunov theorem and the equality above, $A+B_u\bar{\Lambda}+B_a\bar{K}_u$ will be strictly stable if
    \begin{multline*}
    \|SB_u\bar{\Lambda} +(SB_u\bar{\Lambda} )^\mathrm{T}+SB_a(\bar{K}_u-K^\star)\\+(SB_a(\bar{K}_u-K^\star))^\mathrm{T}\|_{\mathrm{F}}<2.
    \end{multline*}
    This holds true if
    \begin{equation}\label{eq:ineq_1f}
    \norm{SB_u}_{\mathrm{F}}\norm{\bar{\Lambda}}_{\mathrm{F}}+\norm{SB_a}_{\mathrm{F}}\norm{\bar{K}_u-K^\star}_{\mathrm{F}}<1.
    \end{equation}
   So we need to obtain a condition that will allow \eqref{eq:ineq_1f} to hold. 
    To that end, note that $\Lambda=0$ and $P_u=P$ from \eqref{eq:are} are feasible solutions to \eqref{eq:opt1}. As a result, by optimality of the limit of the minimizing sequence, we have:
    \begin{align*}
        &\lim_{n\rightarrow\infty}J(\bar{\Lambda}^n,\bar{P}_u^n)\le J(0,P) \Longrightarrow\\&\lim_{n\rightarrow\infty}\norm{K_u(\bar{\Lambda}^n){-}K^\star}_{\mathrm{F}}^2{+} \lim_{n\rightarrow\infty}\mathrm{tr}(\bar{\Lambda}^{n\mathrm{T}}\Gamma\bar{\Lambda}^n)\le \norm{K^\star{-}\bar{K}}_{\mathrm{F}}^2 
    \\&{\Longrightarrow} \norm{\bar{K}_u{-}K^\star}_{\mathrm{F}}^2+ \mathrm{tr}(\bar{\Lambda}^{\mathrm{T}}\Gamma\bar{\Lambda})\le \norm{K^\star{-}\bar{K}}_{\mathrm{F}}^2. 
    \end{align*}
    This implies that
    \begin{equation}\label{eq:ineq_2f}
    \begin{split}        &\norm{\bar{K}_u-K^\star}_{\mathrm{F}}\le \norm{K^\star-\bar{K}}_{\mathrm{F}},\\       &\norm{\bar{\Lambda}}_{\mathrm{F}}\le \frac{1}{\sqrt{\lambda_{\textrm{min}}(\Gamma)}}\norm{K^\star-\bar{K}}_{\mathrm{F}}.
        \end{split}
    \end{equation}
     Hence, using \eqref{eq:ineq_2f}, \eqref{eq:ineq_1f} will hold by imposing the condition
   \begin{multline}\label{eq:cond_stab_proof}
    \hspace{-2mm}\frac{1}{\sqrt{\lambda_{\textrm{min}}(\Gamma)}}\norm{SB_u}_{\mathrm{F}}\norm{K^\star{-}\bar{K}}_{\mathrm{F}}+\norm{SB_a}_{\mathrm{F}}\norm{K^\star{-}\bar{K}}_{\mathrm{F}}<1 \\\Longrightarrow \norm{K^\star-\bar{K}}_{\mathrm{F}}< \frac{\sqrt{\lambda_{\textrm{min}}(\Gamma)}}{\sqrt{\lambda_{\textrm{min}}(\Gamma)}\norm{SB_a}_{\mathrm{F}}+\norm{SB_u}_{\mathrm{F}}}.
    \end{multline}
    Therefore, under \eqref{eq:cond_stab_proof}, $A+B_u\bar{\Lambda}+B_a\bar{K}_u$ is strictly stable. Accordingly, the following LE admits a unique symmetric solution $\bar{P}_u\in\mathbb{R}^{n\times n}$:
    \begin{multline}\label{eq:PulimLE}
    (A+B_u\bar{\Lambda}+B_a\bar{K}_u)^\mathrm{T}\bar{P}_u+\bar{P}_u(A+B_u\bar{\Lambda}+B_a\bar{K}_u)\\+Q-\bar{K}_u^\mathrm{T}R\bar{K}_u=0.
    \end{multline}
    Subtracting \eqref{eq:PulimLE} from \eqref{eq:RicL} for $\Lambda=\bar{\Lambda}^n$ and $P_u=\bar{P}_u^n$, and defining $\tilde{P}_u^n=\bar{P}_u^n-\bar{P}_u$,  $\tilde{\Lambda}_u^n=\bar{\Lambda}^n-\bar{\Lambda}$, $\tilde{K}_u^n=K_u(\bar{\Lambda}^n)-\bar{K}_u$ gives: 
    \begin{align}\nonumber
        (A+&B_u\bar{\Lambda}^n{+}B_aK_u(\bar{\Lambda}^n))^\mathrm{T}\tilde{P}_u^n+\tilde{P}_u^n(A{+}B_u\bar{\Lambda}^n{+}B_aK_u(\bar{\Lambda}^n))\\\nonumber&+\tilde{\Lambda}^{n\mathrm{T}}B_u^\mathrm{T}\bar{P}_u+\bar{P}_uB_u\tilde{\Lambda}^n+\tilde{K}_u^{n\mathrm{T}}B_a^\mathrm{T}\bar{P}_u+\bar{P}_uB_a\tilde{K}_u^n
        \\&-{K}_u^{\mathrm{T}}(\bar{\Lambda}^n)R{K}_u(\bar{\Lambda}^n) +\bar{K}_u^\mathrm{T}R\bar{K}_u=0.\label{eq:tildep}
    \end{align}
    Note that $A_n:=A+B_u\bar{\Lambda}^n+B_aK_u(\bar{\Lambda}^n)$ must remain strictly stable uniformly and it converges to the strictly stable matrix $A_{\infty}:=A+B_u\bar{\Lambda}+B_a\bar{K}_u$ as $n\rightarrow\infty$. In addition, for each $n$ \eqref{eq:tildep} is an LE for $\tilde{P}_u^n$,  in which the last six terms satisfy:
    \begin{multline*}
S_n:=\tilde{\Lambda}^{n\mathrm{T}}B_u^\mathrm{T}\bar{P}_u+\bar{P}_uB_u\tilde{\Lambda}^n+\tilde{K}_u^{n\mathrm{T}}B_a^\mathrm{T}\bar{P}_u+\bar{P}_uB_a\tilde{K}_u^n
        \\-{K}_u^{\mathrm{T}}(\bar{\Lambda}^n)R{K}_u(\bar{\Lambda}^n) +\bar{K}_u^\mathrm{T}R\bar{K}_u\rightarrow0
    \end{multline*}
     as $n\rightarrow\infty$ because $\lim_{n\rightarrow\infty}\bar{\Lambda}^n=\bar{\Lambda}$ and $\lim_{n\rightarrow\infty}K_u(\bar{\Lambda}^n)=\bar{K}_u$. Therefore, by solving \eqref{eq:tildep}:
     \begin{equation*}
    \tilde{P}_u^n=\int_0^{\infty}\mathrm{e}^{A_n^\textrm{T}\tau}S_n\mathrm{e}^{A_n\tau}\textrm{d}\tau\rightarrow\int_0^{\infty}\mathrm{e}^{A_\infty^\textrm{T}\tau}\cdot0\cdot\mathrm{e}^{A_\infty\tau}\textrm{d}\tau=0
     \end{equation*}
     and thus $\bar{P}_u^n\rightarrow\bar{P}_u$ as $n\rightarrow\infty$. In addition, note that since $\bar{P}_u^n$ are positive definite for all $n$, $\bar{P}_u$ must be positive semi-definite, and in fact positive definite owing to Theorem 6.23(ii) in \cite{bacsar1998dynamic}. At the same time, it is straightforward to see that $\bar{K}_u=R^{-1}B_a^\mathrm{T}\bar{P}_u$ since $\bar{K}_u=\lim_{n\rightarrow\infty}K_u(\bar{\Lambda}^n)=\lim_{n\rightarrow\infty}R^{-1}B_a^\mathrm{T}\bar{P}_u^n=R^{-1}B_a^\mathrm{T}\bar{P}_u$, thus $\bar{P}_u$ is the stabilizing solution to \eqref{eq:RicL} for $\Lambda=\bar{\Lambda}$ and hence its minimal one because of Theorem 6.23(vi) in \cite{bacsar1998dynamic}, i.e., $\bar{P}_u\in\mathbb{S}_+$. Thus $(\bar{\Lambda},\bar{P}_u)$ is a feasible solution of \eqref{eq:opt1} that is also minimizing, hence \eqref{eq:opt1} admits a minimizer. The second part follows from the fact that $A+B_u\bar{\Lambda}+B_a\bar{K}_u$ was proved to be strictly stable. \frQED

\textit{Proof of Theorem \ref{th:1}}:
Define the Lagrangian of the optimization \eqref{eq:opt1} as:
\begin{align}\nonumber
\mathcal{L}(\Lambda,P_u,\Pi)=&\norm{R^{-1}B_a^\mathrm{T}P_u-\bar{K}}_{\mathrm{F}}^2+\mathrm{tr}(\Lambda^\mathrm{T}\Gamma\Lambda)\\\nonumber&+\mathrm{tr}\Big(\Pi ((A+B_u\Lambda)^\mathrm{T}P_u+P_u(A+B_u\Lambda)\\&+Q+P_uB_aR^{-1}B_a^\mathrm{T}P_u) \Big),\nonumber
\end{align}
where $\Pi\in\mathbb{R}^{n\times n}$  denotes the Lagrange multiplier. Note that because the left-hand side of the ARE \eqref{eq:RicL} is symmetric, $\Pi$ is symmetric as well.
Applying the first of the three necessary conditions for optimality, namely $\frac{\partial \mathcal{L}}{\partial {\Pi}}=0$, we get
\begin{multline*}
\frac{\partial}{\partial {\Pi}}\mathrm{tr}\Big(\Pi ((A+B_u\Lambda)^\mathrm{T}P_u+P_u(A+B_u\Lambda)\\ \qquad\qquad+Q+P_uB_aR^{-1}B_a^\mathrm{T}P_u) \Big)=0 \stackrel{\eqref{eq:AX}}{\Longrightarrow} \\ (A+B_u\Lambda)^\mathrm{T}P_u{+}P_u(A+B_u\Lambda){+}Q{+}P_uB_aR^{-1}B_a^\mathrm{T}P_u=0,
\end{multline*}
which yields \eqref{eq:eq1}.
Next, by applying the second of the three necessary conditions for optimality, namely $\frac{\partial \mathcal{L}}{\partial {\Lambda}}=0$, and the fact that $\mathrm{tr}(\Lambda^\mathrm{T}\Gamma\Lambda)=\mathrm{tr}(\Gamma\Lambda\Lambda^\mathrm{T})$ we get:
\begin{multline*}
\frac{\partial }{\partial {\Lambda}}
\left(\mathrm{tr}\Big(\Lambda^\mathrm{T}\Gamma\Lambda{+}\Pi ((A+B_u\Lambda)^\mathrm{T}P_u{+}P_u(A+B_u\Lambda)) \right)=0 \\\stackrel{\eqref{eq:AXBXt},\eqref{eq:AXtB},\eqref{eq:AX}}{\Longrightarrow}
2\Gamma\Lambda+2B_u^\mathrm{T}P_u\Pi  =0 \Longrightarrow \Lambda=-\Gamma^{-1}B_u^\mathrm{T}P_u\Pi,
\end{multline*}
where we used the fact that $\Pi$ and $P_u$ are symmetric, hence we obtain \eqref{eq:eq3}.
Finally, applying the third of the three necessary conditions for optimality, namely $\frac{\partial \mathcal{L}}{\partial {P_u}}=0$, and using the fact that $\norm{R^{-1}B_a^\mathrm{T}P_u-\bar{K}}_{\mathrm{F}}^2=\mathrm{tr}((R^{-1}B_a^\mathrm{T}P_u-\bar{K})(R^{-1}B_a^\mathrm{T}P_u-\bar{K})^\mathrm{T})$, we require:
\begin{multline*}
\frac{\partial }{\partial {P_u}}\Big(\mathrm{tr}\Big(R^{-1}B_a^\mathrm{T}P_u^2B_aR^{-1}- R^{-1}B_a^\mathrm{T}P_u\bar{K}^\mathrm{T}\\-\bar{K}P_uB_aR^{-1}\Big)
+\mathrm{tr}\Big(\Pi ((A+B_u\Lambda)^\mathrm{T}P_u\\+P_u(A+B_u\Lambda)+P_uB_aR^{-1}B_a^\mathrm{T}P_u) \Big)=0.
\end{multline*}
Using properties \eqref{eq:AX2B}, \eqref{eq:AXB}, \eqref{eq:AX}, \eqref{eq:AXBX}, we obtain
\begin{multline}\nonumber
(A+B_u\Lambda)\Pi+\Pi(A+B_u\Lambda)^\mathrm{T}+\Pi P_uB_aR^{-1}B_a^\mathrm{T}\\+B_aR^{-1}B_a^\mathrm{T}P_u\Pi-\bar{K}^\mathrm{T}R^{-1}B_a^\mathrm{T}-B_aR^{-1}\bar{K}\\+P_uB_aR^{-2}B_a^\mathrm{T}+B_aR^{-2}B_a^\mathrm{T}P_u=0,
\end{multline}
which, after grouping together the terms containing $\Pi$, is equivalent to \eqref{eq:eq2}. \frQED

\textit{Proof of Lemma \ref{le:grad}}:
Note that we can write the cost as $\tilde{J}(\Lambda)=\tilde{J}_1(\Lambda)+\tilde{J}_2(\Lambda)$, where $\tilde{J}_1(\Lambda)=\mathrm{tr}(\Lambda^\mathrm{T}\Gamma\Lambda)$ and $\tilde{J}_2(\Lambda)=\norm{K_u(\Lambda)-\bar{K}}_{\mathrm{F}}^2$. In that respect, on the one hand, we have:
\begin{equation}\label{eq:dJ1}
\frac{\mathrm{d}\tilde{J}_1(\Lambda)}{\mathrm{d}\Lambda}=\frac{\mathrm{d}}{\mathrm{d}\Lambda}\mathrm{tr}(\Lambda^\mathrm{T}\Gamma\Lambda)=\frac{\mathrm{d}}{\mathrm{d}\Lambda}\mathrm{tr}(\Gamma\Lambda\Lambda^\mathrm{T})\stackrel{\eqref{eq:AXBXt}}{=}2\Gamma\Lambda.
\end{equation}
On the other hand, using the definition of the Frobenius norm we can write $\tilde{J}_2(\Lambda)=\mathrm{tr}((R^{-1}B_a^\mathrm{T}P_u(\Lambda)-\bar{K})(R^{-1}B_a^\mathrm{T}P_u(\Lambda)-\bar{K})^\mathrm{T})$. In that respect, denoting $\Lambda_{ij}=[\Lambda]_{ij}$ and using the commutative property of the derivative operator and the trace, we get:
\begin{align}\nonumber
\frac{\mathrm{d}\tilde{J}_2(\Lambda)}{\mathrm{d}\Lambda_{ij}}&=\frac{\mathrm{d}}{\mathrm{d}\Lambda_{ij}}\Big(\mathrm{tr}\Big(R^{-1}B_a^\mathrm{T}P_u^2B_aR^{-1}- R^{-1}B_a^\mathrm{T}P_u\bar{K}^\mathrm{T}\\\nonumber&\quad-\bar{K}P_uB_aR^{-1}+\bar{K}^\mathrm{T}\bar{K}\Big)\\\nonumber&=\mathrm{tr}\Big(R^{-1}B_a^\mathrm{T}P_u\frac{\mathrm{d}P_u}{\mathrm{d}\Lambda_{ij}}B_aR^{-1}- R^{-1}B_a^\mathrm{T}\frac{\mathrm{d}P_u}{\mathrm{d}\Lambda_{ij}}\bar{K}^\mathrm{T}\\\nonumber&\quad+R^{-1}B_a^\mathrm{T}\frac{\mathrm{d}P_u}{\mathrm{d}\Lambda_{ij}}P_uB_aR^{-1}{-}\bar{K}\frac{\mathrm{d}P_u}{\mathrm{d}\Lambda_{ij}}B_aR^{-1}\Big)\\&\nonumber=\mathrm{tr}\Big(\Big(-\bar{K}^\mathrm{T}R^{-1}B_a^\mathrm{T}-B_aR^{-1}\bar{K}\\&\quad+ P_uB_aR^{-2}B_a^\mathrm{T}{+}B_aR^{-2}B_a^\mathrm{T}P_u          \Big)\frac{\mathrm{d}P_u}{\mathrm{d}\Lambda_{ij}}\Big),\label{eq:dJ2t}
\end{align}
where we also applied the cyclic property of the trace. Next, note that since \eqref{eq:PL} is an equality constraint, we can take the total derivative with respect to $\Lambda_{ij}$ on both sides and obtain:
\begin{align*}
0=&~\frac{\mathrm{d}}{\mathrm{d}\Lambda_{ij}}\Big((A+B_u\Lambda)^\mathrm{T}P_u(\Lambda)+P_u(\Lambda)(A+B_u\Lambda)\\&+Q+P_u(\Lambda)B_aR^{-1}B_a^\mathrm{T}P_u(\Lambda)\Big)\\=&~e_je_i^\mathrm{T}B_u^\mathrm{T}P_u(\Lambda)+(A+B_u\Lambda)^\mathrm{T}\frac{\mathrm{d}P_u}{\mathrm{d}\Lambda_{ij}}+P_u(\Lambda)B_ue_ie_j^\mathrm{T}\\&+\frac{\mathrm{d}P_u}{\mathrm{d}\Lambda_{ij}}(A+B_u\Lambda)+\frac{\mathrm{d}P_u}{\mathrm{d}\Lambda_{ij}}B_aR^{-1}B_a^\mathrm{T}P_u(\Lambda)\\&+P_u(\Lambda)B_aR^{-1}B_a^\mathrm{T}\frac{\mathrm{d}P_u}{\mathrm{d}\Lambda_{ij}},
\end{align*}
where we used the fact that $\frac{\mathrm{d}\Lambda}{\mathrm{d}\Lambda_{ij}}=e_ie_j^\mathrm{T}$.
Grouping together the differential terms yields the LE:
\begin{equation}\label{eq:LE_dPij}
\tilde{A}^\mathrm{T}(\Lambda)\frac{\mathrm{d}P_u}{\mathrm{d}\Lambda_{ij}}+\frac{\mathrm{d}P_u}{\mathrm{d}\Lambda_{ij}}\tilde{A}(\Lambda)+\tilde{Q}(\Lambda)=0
\end{equation}
where
\begin{align}\label{eq:tildeA}
\tilde{A}(\Lambda)&=A+B_u\Lambda+B_aR^{-1}B_a^\mathrm{T}P_u(\Lambda),\\\label{eq:tildeQ}
\tilde{Q}(\Lambda)&=e_je_i^\mathrm{T}B_u^\mathrm{T}P_u(\Lambda)+P_u(\Lambda)B_ue_ie_j^\mathrm{T}.
\end{align}
Note that since $P_u(\Lambda)\in\mathbb{S}_+$, $\tilde{A}(\Lambda)$ is Hurwitz. Therefore, the LE \eqref{eq:LE_dPij} admits the following unique symmetric solution:
\begin{equation}\label{eq:dPij}
\frac{\mathrm{d}P_u}{\mathrm{d}\Lambda_{ij}}=\int_0^\infty \mathrm{e}^{\tilde{A}^\mathrm{T}(\Lambda)\tau}\tilde{Q}(\Lambda)\mathrm{e}^{\tilde{A}(\Lambda)\tau}\mathrm{d}\tau.
\end{equation}
Combining \eqref{eq:dPij} with \eqref{eq:dJ2t}, and using the trace cyclic property, yields
\begin{align}\nonumber
\frac{\mathrm{d}\tilde{J}_2(\Lambda)}{\mathrm{d}\Lambda_{ij}}&=\mathrm{tr}\Big(\Big(-\bar{K}^\mathrm{T}R^{-1}B_a^\mathrm{T}-B_aR^{-1}\bar{K}+ P_uB_aR^{-2}B_a^\mathrm{T}\\\nonumber&\quad+B_aR^{-2}B_a^\mathrm{T}P_u          \Big)\int_0^\infty \mathrm{e}^{\tilde{A}^\mathrm{T}(\Lambda)\tau}\tilde{Q}(\Lambda)\mathrm{e}^{\tilde{A}(\Lambda)\tau}\mathrm{d}\tau\Big)\\\nonumber&=\mathrm{tr}\Big(\tilde{Q}(\Lambda)\int_0^\infty \mathrm{e}^{\tilde{A}(\Lambda)\tau}\Big(-\bar{K}^\mathrm{T}R^{-1}B_a^\mathrm{T}-B_aR^{-1}\bar{K}\\&\quad{+} P_uB_aR^{-2}B_a^\mathrm{T}{+}B_aR^{-2}B_a^\mathrm{T}P_u          \Big)\mathrm{e}^{\tilde{A}^\mathrm{T}(\Lambda)\tau}\mathrm{d}\tau\Big).\hspace{-2mm}\label{eq:dJ2_b}
\end{align}
Note that the integral inside the trace expression above is exactly the solution of the LE \eqref{eq:grad_Pi}. Hence, in light also of the formula \eqref{eq:tildeQ} for $\tilde{Q}$, \eqref{eq:dJ2_b} turns into:
\begin{align*}
\frac{\mathrm{d}\tilde{J}_2(\Lambda)}{\mathrm{d}\Lambda_{ij}}&=\mathrm{tr}\Big((e_je_i^\mathrm{T}B_u^\mathrm{T}P_u(\Lambda)+P_u(\Lambda)B_ue_ie_j^\mathrm{T})\Pi(\Lambda)\Big)\\&=2e_i^\mathrm{T}B_u^\mathrm{T}P_u(\Lambda)\Pi(\Lambda)e_j=2[B_u^\mathrm{T}P_u(\Lambda)\Pi(\Lambda)]_{ij}.
\end{align*}
Therefore 
\begin{equation}\label{eq:dJ2}
\frac{\mathrm{d}\tilde{J}_2(\Lambda)}{\mathrm{d}\Lambda}=2B_u^\mathrm{T}P_u(\Lambda)\Pi(\Lambda).
\end{equation}
Since we defined $\tilde{J}(\Lambda)=\tilde{J}_1(\Lambda)+\tilde{J}_2(\Lambda)$, the results follows by \eqref{eq:dJ1} and \eqref{eq:dJ2}. \frQED

\textit{Proof of Lemma \ref{le:uniformly_bounded}}: 
Based on \eqref{eq:LE_S}, we have:
\begin{align}\nonumber
&(A+B_u\Lambda+B_aK_u(\Lambda))^\mathrm{T}S+S(A+B_u\Lambda+B_aK_u(\Lambda))\\&\nonumber=(A{+}B_aK^\star)^\mathrm{T}S+S(A{+}B_aK^\star)+(K_u(\Lambda){-}K^\star)^\mathrm{T}B_a^\mathrm{T}S\\&\nonumber\quad+SB_a(K_u(\Lambda)-K^\star)+\Lambda B_u^\mathrm{T}S+SB_u\Lambda\\\nonumber&=-2I+(K_u(\Lambda)-K^\star)^\mathrm{T}B_a^\mathrm{T}S+SB_a(K_u(\Lambda)-K^\star)\\&\quad+\Lambda^\mathrm{T}B_u^\mathrm{T}S+SB_u\Lambda.\hspace{-1mm}\label{eq:bounded1}
\end{align}
However:
\begin{multline}\label{eq:bounded2}
\|(K_u(\Lambda)-K^\star)^\mathrm{T}B_a^\mathrm{T}S+SB_a(K_u(\Lambda)-K^\star)\\+\Lambda^\mathrm{T}B_u^\mathrm{T}S+SB_u\Lambda\|_{\mathrm{F}}\\\le 2\norm{SB_a}_{\mathrm{F}}\norm{K_u(\Lambda)-K^\star}_{\mathrm{F}}+2\norm{SB_u}_{\mathrm{F}}\norm{\Lambda}_{\mathrm{F}}\\\le 2\norm{SB_a}_{\mathrm{F}}\norm{K_u(\Lambda)-\bar{K}}_{\mathrm{F}}+2\norm{SB_u}_{\mathrm{F}}\norm{\Lambda}_{\mathrm{F}}\\+2\norm{SB_a}_{\mathrm{F}}\norm{\bar{K}-{K}^\star}_{\mathrm{F}}.
\end{multline}
In addition, if $\Lambda \in\mathcal{S}_0$ then
\begin{align}\nonumber
        &\tilde{J}(\Lambda)\le \tilde{J}(0) \\&\nonumber\Longrightarrow\norm{K_u(\Lambda)-\bar{K}}_{\mathrm{F}}^2+ \mathrm{tr}(\Lambda^{\mathrm{T}}\Gamma\Lambda)\le \norm{K^\star-\bar{K}}_{\mathrm{F}}^2\\&\Longrightarrow \norm{K_u(\Lambda)-\bar{K}}_{\mathrm{F}}^2+\lambda_{\textrm{min}}(\Gamma)\norm{\Lambda}_{\mathrm{F}}^2\le \norm{K^\star-\bar{K}}_{\mathrm{F}}^2.\hspace{-1mm}\label{eq:bounded3}
    \end{align}
    Combining \eqref{eq:bounded2}-\eqref{eq:bounded3} yields
\begin{multline}\nonumber
\hspace{-3mm}\|(K_u(\Lambda)-K^\star)^\mathrm{T}B_a^\mathrm{T}S+SB_a(K_u(\Lambda)-K^\star)+\Lambda^\mathrm{T}B_u^\mathrm{T}S+SB_u\Lambda\|_{\mathrm{F}}\\\le 4(\norm{SB_a}_{\mathrm{F}}+\lambda_{\textrm{min}}^{-1/2}(\Gamma)\norm{SB_u}_{\mathrm{F}})\norm{K^\star-\bar{K}}_{\mathrm{F}}.
\end{multline}
Using \eqref{eq:stab_condition2}, we conclude:
\begin{multline}\label{eq:2-e}
\|(K_u(\Lambda)-K^\star)^\mathrm{T}B_a^\mathrm{T}S+SB_a(K_u(\Lambda)-K^\star)\\+\Lambda^\mathrm{T}B_u^\mathrm{T}S+SB_u\Lambda\|_{\mathrm{F}}\le2\epsilon.
\end{multline}
Combining \eqref{eq:2-e} with \eqref{eq:bounded1}, we obtain
\begin{multline}\label{eq:ineq_unif_stab}
(A+B_u\Lambda+B_aK_u(\Lambda))^\mathrm{T}S\\+S(A+B_u\Lambda+B_aK_u(\Lambda))\preceq -2(1-\epsilon) I.
\end{multline}
Since $S$ is constant, this implies that $A+B_u\Lambda+B_aK_u(\Lambda)$ remains strictly stable uniformly in $\mathcal{S}_0$ and that the energy $\int_0^\infty x^\textrm{T}(t)x(t)\textrm{d}t$ is bounded for every $x_0$ uniformly in $\mathcal{S}_0$.  In light of this and the fact that $K_u(\Lambda)$ is bounded uniformly in $\mathcal{S}_0$, it follows that
the value matrix $P_u(\Lambda)$ is also bounded uniformly in $\mathcal{S}_0$, i.e., $\textrm{sup}_{\Lambda\in\mathcal{S}_0}\norm{P_u(\Lambda)}_\mathrm{F}\le \bar{P}$ for some finite $\bar{P}>0$. Because of this, the cost matrix of the LE \eqref{eq:grad_Pi}, i.e., its last four terms, is uniformly bounded, so $\Pi(\Lambda)$ also remains bounded uniformly, i.e., $\textrm{sup}_{\Lambda\in\mathcal{S}_0}\norm{\Pi(\Lambda)}_\mathrm{F}\le \bar{\Pi}$ for some finite $\bar{\Pi}>0$. \frQED

\textit{Proof of Lemma \ref{le:lipschitz}}: 
Let $\Lambda_1,\Lambda_2\in\mathcal{S}_0$. Denote $\tilde{\Lambda}=\Lambda_1-\Lambda_2$, $\tilde{P_u}=P_u(\Lambda_1)-P_u(\Lambda_2)$, $\tilde{\Pi}=\Pi(\Lambda_1)-\Pi(\Lambda_2)$. Then, using the formula \eqref{eq:grad_formula}:
\begin{align*}
\frac{\mathrm{d}\tilde{J}(\Lambda_1)}{d\Lambda}&-\frac{\mathrm{d}\tilde{J}(\Lambda_2)}{d\Lambda}=2\Gamma(\Lambda_1-\Lambda_2)\\&+2B_u^\mathrm{T}(P_u(\Lambda_1)\Pi(\Lambda_1)-P_u(\Lambda_2)\Pi(\Lambda_2))\\&=2\Gamma\tilde{\Lambda}+2B_u^\mathrm{T}(\tilde{P_u}\Pi(\Lambda_1)+P_u(\Lambda_2)\tilde{\Pi}).
\end{align*}
Using Lemma \ref{le:uniformly_bounded}, this equation implies that:
\begin{multline}\label{eq:dJ1dJ2}
\norm{\frac{\mathrm{d}\tilde{J}(\Lambda_1)}{d\Lambda}-\frac{\mathrm{d}\tilde{J}(\Lambda_2)}{d\Lambda}}_{\mathrm{F}}\le 2\norm{\Gamma}_{\mathrm{F}}\norm{\tilde{\Lambda}}_{\mathrm{F}}\\+2\bar{\Pi}\norm{B_u^\mathrm{T}}_{\mathrm{F}}\norm{\tilde{P}_u}_{\mathrm{F}}+2\bar{P}\norm{B_u^\mathrm{T}}_{\mathrm{F}}\norm{\tilde{\Pi}}_{\mathrm{F}}.
\end{multline}
Hence, to get a Lipschitz continuity result, one needs to relate the errors $\norm{\tilde{P}_u}_{\mathrm{F}}$ and $\norm{\tilde{\Pi}}_{\mathrm{F}}$ back to $\norm{\tilde{\Lambda}}_{\mathrm{F}}$. To that end, recall that $P_u(\Lambda_1)$ and $P_u(\Lambda_2)$ are the minimal positive definite solutions of the AREs:
\begin{align*}\nonumber
&(A+B_u\Lambda_1)^\mathrm{T}P_u(\Lambda_1)+P_u(\Lambda_1)(A+B_u\Lambda_1)\\&\qquad\qquad\qquad\qquad+Q+P_u(\Lambda_1)B_aR^{-1}B_a^\mathrm{T}P_u(\Lambda_1)=0,\\
&(A+B_u\Lambda_2)^\mathrm{T}P_u(\Lambda_2)+P_u(\Lambda_2)(A+B_u\Lambda_2)\\&\qquad\qquad\qquad\qquad+Q+P_u(\Lambda_2)B_aR^{-1}B_a^\mathrm{T}P_u(\Lambda_2)=0.
\end{align*}
Subtracting these equations yields the following quadratic matrix equation for the error $\tilde{P}_u$:
\begin{multline}\label{eq:temp_lips1}
\left(A+B_u\Lambda_1+B_aK_u(\Lambda_1)-\frac{1}{2}B_aR^{-1}B_a^\mathrm{T}\tilde{P}_u\right)^{\mathrm{T}}\tilde{P}_u\\+\tilde{P}_u\left(A+B_u\Lambda_1+B_aK_u(\Lambda_1)-\frac{1}{2}B_aR^{-1}B_a^\mathrm{T}\tilde{P}_u\right)\\
+P_u(\Lambda_2)B_u\tilde{\Lambda}+\tilde{\Lambda}^\mathrm{T}B_u^\mathrm{T}P_u(\Lambda_2)=0.
\end{multline}
Note that this is an equation that yields more than one solutions. Nevertheless, the solution of interest has the property that $\tilde{P}_u\rightarrow0$ as $\tilde{\Lambda}\rightarrow0$. For this solution, since $A+B_u\Lambda_1+B_aK_u(\Lambda_1)$ is Hurwitz as $P_1\in\mathbb{S}_+$, it follows that $\tilde{A}_1(\tilde{P}_u)=A+B_u\Lambda_1+B_aK_u(\Lambda_1)-\frac{1}{2}B_aR^{-1}B_a^\mathrm{T}\tilde{P}_u$ is also Hurwitz for some small $\epsilon>0$ such that $\norm{\tilde{\Lambda}}_{\mathrm{F}}<\epsilon$. Hence, for this solution $\tilde{P}_u$ and for $\norm{\tilde{\Lambda}}_{\mathrm{F}}<\epsilon$, we get from \eqref{eq:temp_lips1} the implicit formula
\begin{equation*}
\tilde{P}_u=\int_0^\infty \mathrm{e}^{\tilde{A}_1^\mathrm{T}(\tilde{P}_u)\tau}(P_u(\Lambda_2)B_u\tilde{\Lambda}+\tilde{\Lambda}^\mathrm{T}B_u^\mathrm{T}P_u(\Lambda_2)) \mathrm{e}^{\tilde{A}_1(\tilde{P}_u)\tau} \mathrm{d}\tau.
\end{equation*}
Therefore, 
\begin{equation}\label{eq:temp_lips2}
\norm{\tilde{P}_u}_{\textrm{F}}\le 2\bar{P}\norm{B_u}_{\textrm{F}}\norm{\tilde{\Lambda}}_{\textrm{F}}\int_0^\infty \norm{\mathrm{e}^{\tilde{A}_1(\tilde{P}_u)\tau}}^2_{\textrm{F}}\mathrm{d}\tau.
\end{equation}
Recall that $\tilde{A}_1(\tilde{P}_u)\rightarrow A+B_u\Lambda_1+B_aK_u(\Lambda_1)$ as $\tilde{\Lambda}\rightarrow0$, and $A+B_u\Lambda_1+B_aK_u(\Lambda_1)$ is strictly stable uniformly for all $\Lambda_1\in\mathcal{S}_0$ owing to Lemma \ref{le:uniformly_bounded}. Hence, the real parts of the eigenvalues of $\tilde{A}_1(\tilde{P}_u)$ remain in the left half plane for some small $\epsilon>0$ such that $\norm{\tilde{\Lambda}}_{\mathrm{F}}<\epsilon$. Hence, for all $\norm{\tilde{\Lambda}}_{\mathrm{F}}<\epsilon$ there exists a finite constant $c_1>0$  such that $\int_0^\infty \norm{\mathrm{e}^{\tilde{A}_1(\tilde{P}_u)\tau}}^2_{\mathrm{F}}\mathrm{d}\tau<c_1$. Putting this in \eqref{eq:temp_lips2} we get the following inequality, locally for $\norm{\tilde{\Lambda}}_{\mathrm{F}}<\epsilon$:
\begin{equation}\label{eq:temp_lips3}
\norm{\tilde{P}_u}_{\mathrm{F}}\le 2c_1\bar{P}\norm{B_u}_{\mathrm{F}}\norm{\tilde{\Lambda}}_{\mathrm{F}}.
\end{equation}
Subsequently, we proceed to derive a similar bound for $\tilde{\Pi}$. Note that subtracting \eqref{eq:grad_Pi} for $\Lambda=\Lambda_1$ and $\Lambda=\Lambda_2$ yields
\begin{multline}\nonumber
(A+B_u\Lambda_1+B_aK_u(\Lambda_1))\tilde{\Pi}+\tilde{\Pi}(A+B_u\Lambda_1{+}B_aK_u(\Lambda_1))^\mathrm{T}\\+\tilde{P_u}B_aR^{-2}B_a^\mathrm{T}+B_aR^{-2}B_a^\mathrm{T}\tilde{P_u}+B_u\tilde{\Lambda}\Pi(\Lambda_2)+\Pi(\Lambda_2)\tilde{\Lambda}^\mathrm{T}B_u^\mathrm{T}\\+B_aR^{-1}B_a^\mathrm{T}\tilde{P}_u\Pi(\Lambda_2)+\Pi(\Lambda_2)\tilde{P}_uB_aR^{-1}B_a^\mathrm{T}=0.\nonumber
\end{multline}
Denote $\tilde{A}(\Lambda_1)=A+B_u\Lambda_1+B_aK_u(\Lambda_1)$ and $\tilde{Q}_1=\tilde{P_u}B_aR^{-2}B_a^\mathrm{T}+B_aR^{-2}B_a^\mathrm{T}\tilde{P_u}+B_u\tilde{\Lambda}\Pi(\Lambda_2)+\Pi(\Lambda_2)\tilde{\Lambda}^\mathrm{T}B_u^\mathrm{T}\\+B_aR^{-1}B_a^\mathrm{T}\tilde{P}_u\Pi(\Lambda_2)+\Pi(\Lambda_2)\tilde{P} _uB_aR^{-1}B_a^\mathrm{T}$. Since $\tilde{A}(\Lambda_1)$ is Hurwitz, we obtain the explicit formula $\tilde{\Pi}=\int_0^\infty \mathrm{e}^{\tilde{A}(\Lambda_1)\tau}\tilde{Q}_1\mathrm{e}^{\tilde{A}^\mathrm{T}(\Lambda_1)\tau}\mathrm{d}\tau$, 
and thus
\begin{equation}\label{eq:temp_lipsPi1}
\norm{\tilde{\Pi}}_{\mathrm{F}}\le \norm{\tilde{Q}_1}_{\mathrm{F}}\int_0^\infty \norm{\mathrm{e}^{\tilde{A}(\Lambda_1)\tau}}_{\mathrm{F}}^2\mathrm{d}\tau\le c_2\norm{\tilde{Q}_1}_{\mathrm{F}},
\end{equation}
where $c_2=\sup_{\Lambda\in\mathcal{S}_0}\int_0^\infty \norm{\mathrm{e}^{\tilde{A}_1(\Lambda)\tau}}_{\mathrm{F}}^2\mathrm{d}\tau>0$ is finite since $\tilde{A}(\Lambda)$ is Hurwitz uniformly in $\mathcal{S}_0$ by Lemma \ref{le:uniformly_bounded}.  
On the other hand, we have 
\begin{multline}\label{eq:temp_lipsPi2}
\norm{\tilde{Q}_1}_{\mathrm{F}}\le 2(\norm{B_aR^{-2}B_a^\mathrm{T}}_\mathrm{F}+\norm{B_aR^{-1}B_a^\mathrm{T}}_\mathrm{F}\bar{\Pi})\norm{\tilde{P}_u}_{\mathrm{F}}\\+2\bar{\Pi}\norm{B_u}_{\mathrm{F}}\norm{\tilde{\Lambda}}_{\mathrm{F}}.
\end{multline}
Combining \eqref{eq:temp_lips3}, \eqref{eq:temp_lipsPi1} and \eqref{eq:temp_lipsPi2}, we conclude
\begin{multline}\label{eq:Ptildeineq}
\norm{\tilde{\Pi}}_{\mathrm{F}}\le c_2\Big(4c_1\bar{P}\norm{B_u}_{\mathrm{F}}\big(\norm{B_aR^{-2}B_a^\mathrm{T}}_\mathrm{F}\\+\norm{B_aR^{-1}B_a^\mathrm{T}}_\mathrm{F}\bar{\Pi}\big)+2\bar{\Pi}\norm{B_u}_{\mathrm{F}}\Big)\norm{\tilde{\Lambda}}_{\mathrm{F}}.
\end{multline}
Therefore, in view of \eqref{eq:dJ1dJ2}, \eqref{eq:temp_lips3} and \eqref{eq:Ptildeineq}, we conclude that $\frac{\mathrm{d}\tilde{J}(\Lambda)}{\mathrm{d}\Lambda}$ is locally Lipschitz on $\mathcal{S}_0$.

Next, note that the set $\mathcal{S}_0$ is bounded; for if $\Lambda\in\mathcal{S}_0$ and $\norm{\Lambda}_{\mathrm{F}}^2>\lambda_{\textrm{max}}(\Gamma)\norm{K^\star-\bar{K}}_\mathrm{F}^2$, then $\tilde{J}(\Lambda)>\tilde{J}(0) $ by definition, and thus $\Lambda\notin\mathcal{S}_0$, which is contradicting. In addition, the set $\mathcal{S}_0$ is closed. To prove this, let $\{\Lambda_n\}_{n\in\mathbb{N}}\in\mathcal{S}_0$ be a sequence such that $\Lambda_n\rightarrow{\Lambda}_{\infty}\in\mathbb{R}^{m_u\times n}$.  Then, based on Lemma \ref{le:uniformly_bounded}, the eigenvalues of the matrix sequence $A+B_u\Lambda+B_aK_u(\Lambda_n)$ remain in the left half plane uniformly on $n\in\mathbb{N}$  while satisfying \eqref{eq:uni_stable}.  Therefore, the sequence of minimal positive definite solutions $P_u(\Lambda_n)$ is bounded, and following \cite{rudin1964principles} one can extract a subsequence $\{\Lambda_{n_k}\}_{k\in\mathbb{N}}$ such that $P_u(\Lambda_{n_k})\rightarrow P_u^\infty$ and $K_u(\Lambda_{n_k})\rightarrow R^{-1}B_a^\textrm{T}P_u^{\infty}=K_u^\infty$  for some positive semi-definite $P_u^\infty\in\mathbb{R}^{n\times n}$. Note that $A+B_u\Lambda_{\infty}+B_aK_u^\infty$ will be strictly stable, because $A+B_u\Lambda_{n_k}+B_aK_u(\Lambda_{n_k})$ are strictly stable uniformly in the sense of \eqref{eq:uni_stable} and converge to $A+B_u\Lambda_\infty+B_aK_u^\infty$. Thus, plugging $\Lambda=\Lambda_{n_k}$ in \eqref{eq:PL} and taking the limit as $k\rightarrow\infty$, we obtain:
\begin{multline}\nonumber
(A+B_u\Lambda_{\infty})^\mathrm{T}P_u^\infty+P_u^\infty(A+B_u\Lambda_{\infty})\\+Q+P_u^\infty B_aR^{-1}B_a^\mathrm{T}P_u^\infty=0,
\end{multline}
meaning that $P(\Lambda_\infty)$ exists and is equal to $P_u^\infty$, and is also in $\mathbb{S}_+$ since $A+B_u\Lambda_{\infty}+B_aK_u^\infty$ was proved to be strictly stable.  Given this, the continuity of $\tilde{J}$ implies  $
\tilde{J}(0)\ge \tilde{J}(\Lambda_{n_k}) \Longrightarrow  \tilde{J}(0)\ge \lim_{k\rightarrow\infty}\tilde{J}(\Lambda_{n_k})= \tilde{J}\left(\lim_{k\rightarrow\infty}\Lambda_{n_k}\right)=\tilde{J}(\Lambda_{\infty}). $
Hence $\Lambda_{\infty}$ is such that $\exists P_{u}(\Lambda_{\infty})\in\mathbb{S}_+$ and $\tilde{J}(0)\ge \tilde{J}(\Lambda_{\infty})$, hence $\Lambda_{\infty}\in\mathcal{S}_0$, which means $\mathcal{S}_0$ is closed. Since $\mathcal{S}_0$ is a finite-dimensional set that is closed and bounded, it is compact. Therefore, since $\frac{\mathrm{d}\tilde{J}(\Lambda)}{\mathrm{d}\Lambda}$ is locally Lipschitz on $\mathcal{S}_0$, it is also globally Lipschitz on $\mathcal{S}_0$. \frQED

\textit{Proof of Theorem \ref{th:convergence}}: 
To prove item 1, it suffices to show that $\Lambda^i\in\mathcal{S}_0$ for all $i\in\mathbb{N}$, because in this case both the ARE \eqref{eq:it1} and the LE \eqref{eq:it2} admit the solutions of interest. We will use induction for that purpose.

- For $i=0$ we have $\Lambda^i\in\mathcal{S}_0$ by design. 

- For a general $i\in\mathbb{N}$, suppose that $\Lambda^i\in\mathcal{S}_0$. We want to show that $\Lambda^{i+1}\in\mathcal{S}_0$. To that end, suppose in seek of contradiction that $\Lambda^{i+1}\notin\mathcal{S}_0$. Define $\hat{\Lambda}(a)=a\Lambda^{i+1}+(1-a)\Lambda^{i}$. Then, there must exist some  $a^\star\in(0,1]$ for which the ARE \eqref{eq:RicL} does not admit a strictly stabilizing solution for $\Lambda=\hat{\Lambda}(a^\star)$, whereas it does for $\Lambda=\hat{\Lambda}(a)$ and for all $a\in[0,a^\star)$. For the latter, it follows that $\tilde{J}(\hat{\Lambda}(a))\le\tilde{J}(\Lambda^i)$ because $\hat{\Lambda}(a)$ is in the direction of the negation of the gradient and $\omega<\frac{2}{L}\frac{\lambda_{\mathrm{min}}(\Gamma^{-1})}{\lambda_{\mathrm{max}}^2(\Gamma^{-1})}$\footnote{This follows from standard argument regarding the convergence of gradient descent; see \cite{bertsekas1997nonlinear}.}. Therefore, since $\Lambda^i\in\mathcal{S}_0$ we obtain $\tilde{J}(\hat{\Lambda}(a))\le\tilde{J}(0)$, and thus $\hat{\Lambda}(a)\in\mathcal{S}_0$ for all $a\in[0,a^\star)$. Using Lemma \ref{le:uniformly_bounded} it follows that $A+B_u\hat{\Lambda}(a)+B_aK_u(\hat{\Lambda}(a))$ remains strictly stable uniformly for all $a\in[0,a^\star)$. Therefore, following continuity arguments similar to those of Lemma \ref{le:lipschitz}, it follows that \eqref{eq:RicL} admits a stabilizing solution also for $\Lambda=\hat{\Lambda}(a^\star)$, which is contradicting. Hence, $\Lambda^{i+1}\in\mathcal{S}_0$. This completes the induction and shows equations \eqref{eq:it1}-\eqref{eq:it3} admit solutions for all $i\in\mathbb{N}$.

To prove items 2-3, note that they follow from i) standard arguments regarding the convergence of gradient descent and the selection of the step size $\omega$ \cite{bertsekas1997nonlinear}; ii)  the fact that the iteration of Algorithm 1 on $\Lambda^i$ is equivalent to \eqref{eq:grad_update}; and iii) the fact that the sequence of iterates $\Lambda^i$ always remains inside $\mathcal{S}_0$ for all $i\in\mathbb{N}$.

To prove item 4, using Lemma \ref{le:grad} we have
\begin{align}\nonumber
\frac{1}{2}\frac{\mathrm{d}^2\tilde{J}(\Lambda)}{\mathrm{d}\Lambda_{ij}\mathrm{d}\Lambda_{kl}}&=\frac{1}{2}\frac{\mathrm{d}}{\mathrm{d}\Lambda_{kl}}\left[\frac{\mathrm{d}\tilde{J}(\Lambda)}{\mathrm{d}\Lambda}\right]_{ij}
\\\nonumber&=\frac{\mathrm{d}}{\mathrm{d}\Lambda_{kl}}e_i^\textrm{T}(\Gamma\Lambda+B_u^\mathrm{T}P_u(\Lambda)\Pi(\Lambda))e_j\\\nonumber&=e_i^\textrm{T}\Gamma\frac{\mathrm{d}\Lambda}{\mathrm{d}\Lambda_{kl}}e_j+e_i^\mathrm{T}B_u^\mathrm{T}\Big(\frac{\mathrm{d}P_u(\Lambda)}{\mathrm{d}\Lambda_{kl}}\Pi(\Lambda)\\&\qquad\qquad\qquad\qquad+P_u(\Lambda)\frac{\mathrm{d}\Pi(\Lambda)}{\mathrm{d}\Lambda_{kl}}\Big)e_j.\label{eq:hessian1}
\end{align}
In addition, $e_i^\textrm{T}\Gamma\frac{\mathrm{d}\Lambda}{\mathrm{d}\Lambda_{kl}}e_j=e_i^\textrm{T}\Gamma e_ke_l^\textrm{T}e_j=[\Gamma]_{ik}\delta_{l,j}$. Thus, \eqref{eq:hessian1} becomes
\begin{multline}\label{eq:hessian2}
\frac{1}{2}\frac{\mathrm{d}^2\tilde{J}(\Lambda)}{\mathrm{d}\Lambda_{ij}\mathrm{d}\Lambda_{kl}}=[\Gamma]_{ik}\delta_{l,j}\\+e_i^\mathrm{T}B_u^\mathrm{T}\Big(\frac{\mathrm{d}P_u(\Lambda)}{\mathrm{d}\Lambda_{kl}}\Pi(\Lambda)+P_u(\Lambda)\frac{\mathrm{d}\Pi(\Lambda)}{\mathrm{d}\Lambda_{kl}}\Big)e_j.
\end{multline}
Using the bounds of Lemma \ref{le:uniformly_bounded}, which are independent of $\Gamma$, we get from \eqref{eq:hessian2} the inequality:
\begin{multline}\label{eq:hessian3}
\frac{1}{2}\frac{\mathrm{d}^2\tilde{J}(\Lambda)}{\mathrm{d}\Lambda_{ij}\mathrm{d}\Lambda_{kl}}\ge[\Gamma]_{ik}\delta_{l,j}
\\-\norm{B_u^\mathrm{T}}_{\mathrm{F}}\Big(\bar{\Pi}\norm{\frac{\mathrm{d}P_u(\Lambda)}{\mathrm{d}\Lambda_{kl}}}_{\mathrm{F}}+\bar{P}\norm{\frac{\mathrm{d}\Pi(\Lambda)}{\mathrm{d}\Lambda_{kl}}}_{\mathrm{F}}\Big).
\end{multline}

The equation above implies the possibility that if $\Gamma$ is sufficiently large in terms of eigenvalues, then $\tilde{J}$ is strictly convex on $\mathcal{S}_0$. However, it first needs to be shown that $\norm{\frac{\mathrm{d}P_u(\Lambda)}{\mathrm{d}\Lambda_{kl}}}_{\mathrm{F}}$ and $\norm{\frac{\mathrm{d}\Pi(\Lambda)}{\mathrm{d}\Lambda_{kl}}}_{\mathrm{F}}$ remain bounded uniformly as $\lambda_{\textrm{min}}(\Gamma)$ increases. To that end, from \eqref{eq:dPij}, we get
\begin{equation}\nonumber
\norm{\frac{\mathrm{d}P_u}{\mathrm{d}\Lambda_{kl}}}_{\mathrm{F}}{\le}2\norm{B_u}_{\mathrm{F}}\bar{P}\int_0^\infty\norm{\mathrm{e}^{(A+B_u\Lambda+B_aR^{-1}B_a^\mathrm{T}P_u(\Lambda))\tau}}^2_{\mathrm{F}}\mathrm{d}\tau.
\end{equation}
But by Lemma \ref{le:uniformly_bounded}, $A+B_u\Lambda+B_aR^{-1}B_a^\mathrm{T}P_u(\Lambda)$ is strictly stable uniformly on $\mathcal{S}_0$ and independently of $\Gamma$ (see \eqref{eq:uni_stable}), so there exists a finite constant $c_1$ independent of $\Gamma$ and $\Lambda$ such that $\int_0^\infty\norm{\mathrm{e}^{(A+B_u\Lambda+B_aR^{-1}B_a^\mathrm{T}P_u(\Lambda))\tau}}_{\mathrm{F}}^2\mathrm{d}\tau<c_1$, and thus
\begin{equation}\label{eq:DPbound}
\norm{\frac{\mathrm{d}P_u(\Lambda)}{\mathrm{d}\Lambda_{kl}}}_{\mathrm{F}}\le2\norm{B_u}_{\mathrm{F}}\bar{P}c_1.
\end{equation}

On the other hand, a differentiation of \eqref{eq:grad_Pi} yields:
\begin{align*}\nonumber
&(A+B_u\Lambda+B_aR^{-1}B_a^\mathrm{T}P_u(\Lambda))\frac{\mathrm{d}\Pi(\Lambda)}{\mathrm{d}\Lambda_{kl}}\\&+\frac{\mathrm{d}\Pi(\Lambda)}{\mathrm{d}\Lambda_{kl}}(A+B_u\Lambda{+}B_aR^{-1}B_a^\mathrm{T}P_u(\Lambda))^\mathrm{T}\\&+\frac{\mathrm{d}P_u(\Lambda)}{\mathrm{d}\Lambda_{kl}}B_aR^{-2}B_a^\mathrm{T}+B_aR^{-2}B_a^\mathrm{T}\frac{\mathrm{d}P_u(\Lambda)}{\mathrm{d}\Lambda_{kl}}\\&+B_ue_ie_j^\textrm{T}\Pi(\Lambda)+\Pi(\Lambda)e_je_i^\textrm{T}B_u^\textrm{T}
\\&+B_aR^{-1}B_a^\mathrm{T}\frac{\mathrm{d}P_u(\Lambda)}{\mathrm{d}\Lambda_{kl}}\Pi(\Lambda)+\Pi(\Lambda)\frac{\mathrm{d}P_u(\Lambda)}{\mathrm{d}\Lambda_{kl}}B_aR^{-1}B_a^\mathrm{T}=0.
\end{align*}
This is an LE for $\frac{\mathrm{d}\Pi(\Lambda)}{\mathrm{d}\Lambda_{kl}}$, the plant matrix of which, i.e., $A+B_u\Lambda+B_aR^{-1}B_a^\mathrm{T}P_u(\Lambda)$, is uniformly strictly stable owing to Lemma \ref{le:uniformly_bounded}, and the cost matrix of which, i.e., its last six terms, is uniformly bounded owing to Lemma \ref{le:uniformly_bounded} and \eqref{eq:DPbound}. Hence, following the line of proof of the previous paragraph, there exists a finite constant $c_2>0$ such that for all $\Lambda\in\mathcal{S}_0$:
\begin{equation}\label{eq:DPibound}
\norm{\frac{\mathrm{d}\Pi(\Lambda)}{\mathrm{d}\Lambda_{kl}}}_{\mathrm{F}}\le c_2.
\end{equation}
Combining \eqref{eq:hessian3} with \eqref{eq:DPbound}-\eqref{eq:DPibound} we obtain
\begin{multline}\label{eq:hessianineq1}
\frac{1}{2}\frac{\mathrm{d}^2\tilde{J}(\Lambda)}{\mathrm{d}\Lambda_{ij}\mathrm{d}\Lambda_{kl}}\ge[\Gamma]_{ik}\delta_{l,j}
\\-\norm{B_u^\mathrm{T}}_{\mathrm{F}}\Big(2c_1\norm{B_u}_{\mathrm{F}}\bar{P}\bar{\Pi}+\bar{P}c_2\Big).
\end{multline}
To proceed, define the Hessian matrix:
\begin{equation*}
\frac{\mathrm{d}^2\tilde{J}(\Lambda)}{\mathrm{d}\textrm{vec}(\Lambda)^2}=\begin{bmatrix} \frac{\mathrm{d}^2\tilde{J}(\Lambda)}{\mathrm{d}\Lambda_{11}^2} & \frac{\mathrm{d}^2\tilde{J}(\Lambda)}{\mathrm{d}\Lambda_{11}\mathrm{d}\Lambda_{21}} & \ldots &  \frac{\mathrm{d}^2\tilde{J}(\Lambda)}{\mathrm{d}\Lambda_{11}\mathrm{d}\Lambda_{m_u,n}} \\ \frac{\mathrm{d}^2\tilde{J}(\Lambda)}{\mathrm{d}\Lambda_{21}\mathrm{d}\Lambda_{11}} & \frac{\mathrm{d}^2\tilde{J}(\Lambda)}{\mathrm{d}\Lambda_{21}^2} & \ldots   & \frac{\mathrm{d}^2\tilde{J}(\Lambda)}{\mathrm{d}\Lambda_{21}\mathrm{d}\Lambda_{m_u,n}} \\\vdots & \vdots & \ddots & \vdots \\ \frac{\mathrm{d}^2\tilde{J}(\Lambda)}{\mathrm{d}\Lambda_{m_u,n}\mathrm{d}\Lambda_{11}} & \frac{\mathrm{d}^2\tilde{J}(\Lambda)}{\mathrm{d}\Lambda_{m_u,n}\mathrm{d}\Lambda_{21}} & \ldots  & \frac{\mathrm{d}^2\tilde{J}(\Lambda)}{\mathrm{d}\Lambda_{m_u,n}^2} \end{bmatrix}.
\end{equation*}
Based on the inequality \eqref{eq:hessianineq1}, this Hessian becomes diagonally dominant for sufficiently large values of $[\Gamma]_{ii}$, $i=1,\ldots,m_u$, while keeping $[\Gamma]_{ij}$ constant for $i\ne j$, $i,j=1,\ldots,m_u$. Therefore, there exists $\gamma^\star>0$ such that if $\Gamma\succ\gamma^\star I$ then $\frac{\mathrm{d}^2\tilde{J}(\Lambda)}{\mathrm{d}\textrm{vec}(\Lambda)^2}$ is positive definite uniformly. Therefore, if $\Gamma\succ\gamma^\star I$ then $\tilde{J}$, when viewed as a function of $\textrm{vec}(\Lambda)$, is strictly convex, and thus any stationary point is a minimum. Therefore, by item 3, Algorithm $1$ converges to a minimum of $\tilde{J}$, and thus of \eqref{eq:opt1}. \frQED

\textit{Proof of Proposition \ref{pr:1}:} 
Since $\mathrm{Ran}(B_a)\subseteq\mathrm{Ran}(B_u)$, there exists a matrix $E\in\mathbb{R}^{m_u\times m_a}$ such that $B_a=B_uE$. In light of this, \eqref{eq:RicL} becomes $(A+B_u{\Lambda})^\mathrm{T}P_u+P_u(A+B_u{\Lambda})+Q+P_uB_uER^{-1}E^\textrm{T}B_u^\mathrm{T}P_u=0 $.
Choosing $\Lambda=\bar{\Lambda}=-\frac{1}{2}R_u^{-1}B_u^\mathrm{T}P_u$ for some positive definite matrix $R_u\succ0$ to be selected, this equation becomes
\begin{equation}\label{eq:temprob1}
    A^\mathrm{T}P_u+P_uA+Q+P_uB_u(ER^{-1}E^\textrm{T}-R_u^{-1})B_u^\mathrm{T}P_u=0.
\end{equation}
Selecting $ R_u=(aI+ER^{-1}E^\textrm{T})^{-1}$ for some positive $a>0$, we have $R_u^{-1}=aI+ER^{-1}E^\textrm{T}$. Therefore, \eqref{eq:temprob1} yields
\begin{equation}\label{eq:temprob2}
    A^\mathrm{T}P_u+P_uA+Q-aP_uB_uB_u^\mathrm{T}P_u=0.
\end{equation}
Since $A$ is Hurwitz, we can define a constant matrix $V$ as the unique positive definite solution of the LE
\begin{equation}\label{eq:tempV}
    A^\mathrm{T}V+VA+Q=0.
\end{equation}
Subtracting \eqref{eq:temprob2} from \eqref{eq:tempV}, we obtain $A^\mathrm{T}(V-P_u)+(V-P_u)A+aP_uB_uB_u^\mathrm{T}P_u=0$. Since $A$ is Hurwitz and $aP_uB_uB_u^\mathrm{T}P_u\succeq0$, we obtain $P_u\preceq V$ by the Lyapunov theorem. Since $P_u$ is positive semidefinite, this implies $\norm{P_u}_\mathrm{F}\le \norm{V}_\mathrm{F}$. Next, by applying norms to \eqref{eq:temprob2}:
\begin{multline*}
a\norm{P_uB_uB_u^\mathrm{T}P_u}_\mathrm{F}=\norm{A^\mathrm{T}P_u+P_uA+Q}_{\mathrm{F}}\\\le 2\norm{A}_\mathrm{F}\norm{P_u}_\mathrm{F}+\norm{Q}_\mathrm{F}\le 2\norm{A}_\mathrm{F}\norm{V}_\mathrm{F}+\norm{Q}_\mathrm{F}.
\end{multline*}
Since $A, V, Q$ are constant matrices, as $a\rightarrow\infty$ we have $\norm{P_uB_uB_u^\mathrm{T}P_u}_\mathrm{F}\rightarrow0$, hence $\norm{B_u^\mathrm{T}P_u}_\mathrm{F}\rightarrow0$, and hence,
$\norm{K_u(\bar{\Lambda})}_\mathrm{F}=\norm{R^{-1}B_a^\mathrm{T}P_u(\bar{\Lambda})}_\mathrm{F}=\norm{R^{-1}E^\mathrm{T}B_u^\mathrm{T}P_u(\bar{\Lambda})}_\mathrm{F}\rightarrow0$. Therefore, for every $\epsilon>0$ there exists $\Lambda\in\mathbb{R}^{m_u\times n}$ such that $\norm{K_u(\Lambda)}_{\mathrm{F}}^2<\frac{\epsilon}{2}$. In addition, for this $\Lambda$, one can find $\gamma_1>0$ such that if $\Gamma\prec\gamma_1I$ then $\textrm{tr}(\Lambda^\textrm{T}\Gamma\Lambda)<\frac{\epsilon}{2}$, and hence $\tilde{J}(\Lambda)<\epsilon$ in total. Summarizing, we conclude for every $\epsilon>0$ there exists $\gamma_1>0$ such that if $\Gamma\prec\gamma_1 I$ then $\inf_{\Lambda} \tilde{J}(\Lambda)<\epsilon$.

Next,   let $\{{\Lambda}^n,{P}_u^n\}_{n\in\mathbb{N}}$ be a minimizing sequence of \eqref{eq:opt1}, and $K^n_u=R^{-1}B_a^\textrm{T}P_u^n$. From the constraints of \eqref{eq:opt1} we get:
\begin{multline}\label{eq:tempabove}
    (A+B_u\Lambda^n)^\mathrm{T}P_u^n+P_u^n(A+B_u\Lambda^n)+Q\\+K_u^{n\textrm{T}}RK_u^n=0.
\end{multline} 
Since $P_u^n$ is constrained to be positive definite and $Q\succ0$, it follows from \eqref{eq:tempabove} that $A+B_u\Lambda^n$ is strictly stable for all $n\in\mathbb{N}$. Assume now, in seek for contradiction, that $\sup_{n\in\mathbb{N}}\alpha(A+B_u\Lambda^n)=0$. Then, for every $a>0$ there exists $n^\star>0$ such that if $n\ge n^\star$ then $\alpha(A+B_u\Lambda^n)\ge-a$. In addition, if we choose $a$ sufficiently close to $0$, then this spectral abscissa corresponds to an (without loss of generality simple) eigenvalue, denoted as $\lambda_n$, that is not controllable by $B_a$; otherwise, $A+B_u\Lambda^n+B_aK_u^n$ would be unstable for some finite $n$ (because $K_u^n$ is the optimal attack gain against $A+B_u\Lambda^n$, and $A+B_u\Lambda^n$ is bounded), which contradicts $P_u^n\in\mathbb{S}_+$.  Given this, denote the uncontrollable eigenvector of $(A+B_u\Lambda^n, B_a)$ as $v_n$, and let us decompose $\Lambda^n=\Lambda^nv_nv_n^\dagger+\Lambda^n(I-v_nv_n^\dagger )$. Note, $\Lambda^nv_nv_n^\dagger\ne0$ uniformly; otherwise, the eigenvalue $\lambda^n$ corresponding to $v_n$ would not tend to $0$ as $n\rightarrow\infty$, since it is not controllable by $B_a$. Next, define $\hat{\Lambda}^n=\Lambda^n(I-v_n v_n^\dagger)$, which satisfies $\textrm{tr}(\hat{\Lambda}^{n\textrm{T}}\Gamma\hat{\Lambda}^n)<\textrm{tr}({\Lambda}^{n\textrm{T}}\Gamma{\Lambda}^n)$ uniformly. Since the difference between $\hat{\Lambda}^n$ and ${\Lambda}^n$ is only over the space that is uncontrollable by $B_a$, it follows that $K_u(\hat{\Lambda}^n)=K_u(\Lambda^n)$. Therefore, $\tilde{J}(\hat{\Lambda}^n)<\tilde{J}({\Lambda}^n)$ uniformly, i.e., ${\Lambda}^n$ is not a minimizing sequence. Therefore, we arrive at a contradiction, and $\sup_{n\in\mathbb{N}}\alpha(A+B_u\Lambda^n)<0$.

Matrices $\Lambda^n$, $K_u^n$ are uniformly bounded because they are a minimizing sequence of \eqref{eq:opt1}, so there exist subsequences of $\Lambda^{n_k}$, $K_u^{n_k}$, $P_u^{n_k}$ of $\Lambda^{n}$, $K_u^{n}$, $P_u^n$ such that $\Lambda^{n_k}\rightarrow\bar{\Lambda}, K_u^{n_k}\rightarrow\bar{K}_u$ for some bounded limits $\bar{\Lambda},~\bar{K}_u$. Since $\sup_{n\in\mathbb{N}}\alpha(A+B_u\Lambda^n)<0$, $A+B_u\bar{\Lambda}$ is strictly stable, hence there is a strictly positive definite matrix $\bar{P}_u\succ0$ such that
\begin{equation}\label{eq:subtractalt}
(A+B_u\bar{\Lambda})^\mathrm{T}\bar{P}_u+\bar{P}_u(A+B_u\bar{\Lambda})+Q+\bar{K}_u^\textrm{T}R\bar{K}_u=0.
\end{equation}
Defining $\tilde{P}_u^{n_k}=P_u^{n_k}-\bar{P}_u$, $\tilde{\Lambda}^{n_k}=\Lambda^{n_k}-\bar{\Lambda}$, subtracting \eqref{eq:subtractalt} from \eqref{eq:tempabove} for the subsequence we get
\begin{multline}\label{eq:subtractalt2}
(A+B_u\bar{\Lambda})^\mathrm{T}\tilde{P}_u^{{n_k}}+\tilde{P}_u^{{n_k}}(A+B_u\bar{\Lambda})-\bar{K}_u^\textrm{T}R\bar{K}_u\\+K_u^{{n_k}\textrm{T}}RK_u^{n_k}+P_u^{n_k}B_u\tilde{\Lambda}^{n_k}+\tilde{\Lambda}^{\textrm{T}{n_k}}B_u^\textrm{T}P_u^{n_k}=0.
\end{multline}
Note, $K_u^{n_k}\rightarrow\bar{K}_u$ and $\tilde{\Lambda}^{n_k}\rightarrow0$. Moreover, $P_u^{n_k}$ is bounded uniformly because  $A+B_u\Lambda^{n_k}$ is bounded and strictly stable uniformly and because $Q+K_u^{n\textrm{T}}RK_u^n$ is bounded uniformly. Hence, \eqref{eq:subtractalt2} implies $\tilde{P}_u^{n_k}\rightarrow0$ and thus $P_u^{n_k}\rightarrow\bar{P}_u$. Finally, since this implies $\bar{K}_u=R^{-1}B_a^\textrm{T}\bar{P}_u,$ rearranging \eqref{eq:subtractalt} yields
\begin{multline}\nonumber
(A+B_u\bar{\Lambda}+B_a\bar{K}_u)^\mathrm{T}\bar{P}_u+\bar{P}_u(A+B_u\bar{\Lambda}+B_a\bar{K}_u)\\+Q-\bar{K}_u^\textrm{T}R\bar{K}_u=0.
\end{multline}
From the first paragraph, for every $\epsilon$ there is $\gamma^\star>0$ such that if $\Gamma\prec\gamma^\star I$ then $\norm{\bar{K}_u}_{\mathrm{F}}<\epsilon$ and hence $Q-\bar{K}_u^\textrm{T}R\bar{K}_u\succ0$ for small enough $\epsilon$, so the above equation implies $A+B_u\bar{\Lambda}+B_a\bar{K}_u$ is strictly stable.  Thus $(\bar{\Lambda},\bar{P}_u)$ is a feasible solution of \eqref{eq:opt1} that is also minimizing, hence \eqref{eq:opt1} admits a minimizer.

\textit{Proof of Proposition \ref{pr:2}:} Let $\Lambda$ be in the domain of $\hat{J}$ and $\tilde{J}$. Using the value of the optimal control problem \eqref{eq:Jper} and the assumption $\hat{Q}\preceq Q$, $\hat{R}\succeq R$, we have
\begin{multline*}
x_0^\mathrm{T}\hat{P}_u(\Lambda)x_0=\max_K\int_0^\infty(x^\mathrm{T}(\tau)\hat{Q}x(\tau)-a^\mathrm{T}(\tau)\hat{R}a(\tau))\mathrm{d}\tau\\\le  \max_K\int_0^\infty(x^\mathrm{T}(\tau){Q}x(\tau)-a^\mathrm{T}(\tau){R}a(\tau))\mathrm{d}\tau=x_0^\mathrm{T}{P}_u(\Lambda)x_0.
\end{multline*}
Since this holds for every $x_0$, it follows  $\hat{P}_u(\Lambda)\preceq P_u(\Lambda)$. Accordingly, by combining this with the fact that $\hat{R}\succeq R$:
\begin{multline*}
\norm{\hat{R}^{-1}B_a^\mathrm{T}\hat{P}_u(\Lambda)}_\textrm{F}^2\le\norm{{R}^{-1}B_a^\mathrm{T}{P}_u(\Lambda)}_\textrm{F}^2 \\ \Longrightarrow \norm{\hat{K}_u(\Lambda)-\bar{K}}_\textrm{F}^2\le\norm{{K}_u(\Lambda)-\bar{K}}_\textrm{F}^2  \Longrightarrow \hat{J}(\Lambda)\le\tilde{J}(\Lambda),
\end{multline*}
which yields $\hat{J}(\Lambda^\star)\le \tilde{J}(\Lambda^\star)$ and proves item 1.

To prove item 2, there exists $\lambda\ge1$ so that $\hat{R}=\frac{1}{\lambda} \bar{R}$, $\hat{Q}=\lambda \bar{Q}$,  $\bar{R}\succ R$, $\bar{Q}\prec Q$. Given this, let $\bar{P}_u(\Lambda^\star)\in\mathbb{S}_+$ solve
\begin{multline}\label{eq:pert1}
(A+B_u{\Lambda}^\star)^\mathrm{T}\bar{P}_u(\Lambda^\star)+\bar{P}_u(\Lambda^\star)(A+B_u{\Lambda}^\star)\\+\bar{Q}+\bar{P}_u(\Lambda^\star)B_a\bar{R}^{-1}B_a^\mathrm{T}\bar{P}_u(\Lambda^\star)=0,
\end{multline}
and denote as $\bar{J}(\Lambda^\star)$ the cost with matrices $\bar{Q}$, $\bar{R}$.

In Proposition \ref{pr:1} it was shown that for every $\epsilon>0$ there exists $\gamma>0$ such that if $\Gamma\prec \gamma I$ then $\tilde{J}(\Lambda^\star)<\epsilon$. In item 1, we also proved that $\bar{J}(\Lambda^\star)\le\tilde{J}(\Lambda^\star)$. Combining these, for every $\epsilon>0$ there exists $\gamma>0$ such that if $\Gamma\prec \gamma I$ then $\bar{J}(\Lambda^\star)<\epsilon$. Hence, $B_a^\textrm{T}\bar{P}_u(\Lambda^\star)\rightarrow0$ and thus $\bar{P}_u(\Lambda^\star)B_a\bar{R}^{-1}B_a^\textrm{T}\bar{P}_u(\Lambda^\star)\rightarrow0$ as $\gamma\rightarrow0$. Moreover, note that from \eqref{eq:Phat}:
\begin{multline}\label{eq:pert2}
    (A+B_u\Lambda^\star)^\mathrm{T}\hat{P}_u(\Lambda^\star)+\hat{P}_u(\Lambda^\star)(A+B_u\Lambda^\star)\\+\lambda(\bar{Q}+\hat{P}_u(\Lambda^\star)B_a\bar{R}^{-1}B_a^\mathrm{T}\hat{P}_u(\Lambda^\star))=0.
\end{multline}
Comparing \eqref{eq:pert1}-\eqref{eq:pert2}, since $\bar{P}_u(\Lambda^\star)B_a\bar{R}^{-1}B_a^\textrm{T}\bar{P}_u(\Lambda^\star)\rightarrow0$ as $\gamma\rightarrow0$, it follows that $\hat{P}_u(\Lambda^\star)\approx \lambda\bar{P}_u(\Lambda^\star)$, where the approximation becomes exact as $\gamma\rightarrow0$. Hence, as $\gamma\rightarrow0$, $B_a^\textrm{T}\hat{P}_u(\Lambda^\star)\rightarrow\lambda B_a^\textrm{T}\bar{P}_u(\Lambda^\star)\rightarrow0$ and thus $\hat{K}_u(\Lambda^\star)\rightarrow0$.   Therefore, for every $\epsilon>0$ there exists $\gamma^\star>0$ such that if $\Gamma\prec \gamma^\star I$ then $\norm{\hat{K}_u(\Lambda^\star)}_\textrm{F}^2=\norm{\hat{R}^{-1}B_a^\textrm{T}\hat{P}_u(\Lambda^\star)}_\textrm{F}^2<\frac{\epsilon}{2}$, and thus also $\hat{J}(\Lambda^\star)<\epsilon$, which yields the required result. \frQED

\end{document}